\documentclass[useAMS,usenatbib]{mn2e}

\usepackage{latexsym,graphicx,natbib}

\usepackage{color}

\newcommand\kms{{\rm\,km\,s^{-1}}}
\newcommand\msun{\rm\,M_\odot}
\newcommand\rsun{\rm\,R_\odot}
\newcommand\lsun{\rm\,L_\odot}
\newcommand\hii{H\,{\sc ii} \,}

\newcommand\myr{\msun \, {\rm yr}^{-1}}
\DeclareRobustCommand{\ion}[2]{%
\relax\ifmmode
\ifx\testbx\f
{\mathrm{#1\,\textsc{#2}}}\else {\mathrm{#1\,\mathsc{#2}}}\fi
\else\textup{#1\,{\mdseries\textsc{#2}}}%
\fi}
\newcommand{\MC}{\multicolumn}

\def\apgt{\ {\raise-.5ex\hbox{$\buildrel>\over\sim$}}\ }
\def\aplt{\ {\raise-.5ex\hbox{$\buildrel<\over\sim$}}\ }

\title[MN18 and its bipolar nebula]{The blue supergiant MN18 and its bipolar circumstellar nebula\footnotemark[0]\thanks{Based
on observations made with the Southern African Large Telescope,
programs \mbox{2011-3-RSA\_OTH-002} and
\mbox{2013-1-RSA\_OTH-014}, and the 2.2m MPG telescope at La Silla
(ESO), program ID P092.A-9020.}}
\author[V.V.Gvaramadze et al.]
       {V. V.~Gvaramadze,$^{1,2,3}$\thanks{E-mail: vgvaram@mx.iki.rssi.ru} A. Y.~Kniazev,$^{1,4,5}$ J. M.~Bestenlehner,$^{6,7}$
       J.~Bodensteiner,$^{8,9}$
        \newauthor
        N.~Langer,$^{6}$ J.~Greiner,$^{9}$ E. K.~Grebel,$^{10}$ L. N.~Berdnikov$^{1,3,11}$ and Y.~Beletsky$^{12}$ \\
        $^{1}$Sternberg Astronomical Institute, Lomonosov Moscow State University, Universitetskij Pr. 13, Moscow 119992, Russia \\
        $^{2}$Space Research Institute, Russian Academy of Sciences, Profsoyuznaya 84/32, 117997 Moscow, Russia \\
        $^{3}$Isaac Newton Institute of Chile, Moscow Branch, Universitetskij Pr. 13, Moscow 119992, Russia \\
        $^{4}$South African Astronomical Observatory, PO Box 9, 7935 Observatory, Cape Town, South Africa \\
        $^{5}$Southern African Large Telescope Foundation, PO Box 9, 7935 Observatory, Cape Town, South Africa \\
        $^{6}$Argelander-Institut f\"ur Astronomie der Universit\"at Bonn, Auf dem H\"ugel 71, 53121, Bonn, Germany \\
        $^{7}$Max-Planck Institute for Astronomy, D-69117, Heidelberg, Germany \\
        $^{8}$Technische Universit\"at M\"unchen, Physik Dept., James-Franck-Str., 85748 Garching, Germany \\
        $^{9}$Max-Planck-Institut f\"ur extraterrestrische Physik, Giessenbachstra{\ss}e 1, D-85748, Garching, Germany \\
        $^{10}$Astronomisches Rechen-Institut, Zentrum f\"ur Astronomie der Universit\"at Heidelberg, M\"onchhofstr.\ 12--14, 69120 Heidelberg, Germany \\
        $^{11}$Astronomy and Astrophysics Research division, Entoto Observatory and Research Center, P.O.Box 8412, Addis Ababa, Ethiopia \\
        $^{12}$Las Campanas Observatory, Carnegie Institution of Washington, Colina el Pino, Casilla 601 La Serena, Chile \\
        }
\begin{document}

\date{Accepted 2015 August 25. Received 2015 August 24; in original form 2015 July 7}


\maketitle

\label{firstpage}

\begin{abstract}
We report the results of spectrophotometric observations of the
massive star MN18 revealed via discovery of a bipolar nebula
around it with the {\it Spitzer Space Telescope}. Using the
optical spectrum obtained with the Southern African Large
Telescope, we classify this star as B1\,Ia. The evolved status of
MN18 is supported by the detection of nitrogen overabundance in
the nebula, which implies that it is composed of processed
material ejected by the star. We analysed the spectrum of MN18 by
using the code {\sc cmfgen}, obtaining a stellar effective
temperature of $\approx$21 kK. The star is highly reddened,
$E(B-V)$$\approx$2 mag. Adopting an absolute visual magnitude of
$M_V$=$-$6.8$\pm$0.5 (typical of B1 supergiants), MN18 has a
luminosity of $\log L/\lsun$$\approx$5.42$\pm$0.30, a mass-loss
rate of $\approx$$(2.8-4.5)\times10^{-7} \, \myr$, and resides at
a distance of $\approx$5.6$^{+1.5} _{-1.2}$ kpc. We discuss the
origin of the nebula around MN18 and compare it with similar
nebulae produced by other blue supergiants in the Galaxy
(Sher\,25, HD\,168625, [SBW2007]\,1) and the Large Magellanic
Cloud (Sk$-$69$\degr$202). The nitrogen abundances in these
nebulae imply that blue supergiants can produce them from the main
sequence stage up to the pre-supernova stage. We also present a
$K$-band spectrum of the candidate luminous blue variable MN56
(encircled by a ring-like nebula) and report the discovery of an
OB star at $\approx$17 arcsec from MN18. The possible membership
of MN18 and the OB star of the star cluster Lynga\,3 is discussed.
\end{abstract}


\section{Introduction}
\label{sec:int}

Massive stars change their mass-loss rate, and the velocity,
symmetry and composition of their copious winds in the course of
their evolution. The mass loss along with stellar rotation and
magnetic fields, and effects of binarity (various modes of mass
transfer between binary components, mergers of the components,
common-envelope ejection) result in the formation of different
types of circumstellar nebulae, of which the hourglass-like type
is most spectacular. The first and the best known example of such
a nebula associated with a massive star is that created by the
Large Magellanic Cloud B-type supergiant Sk$-$69$\degr$202 -- the
progenitor of the SN\,1987A (Plait et al. 1995; Burrows et al.
1995). Later, a number of Galactic analogs of this nebula were
discovered around the already known blue supergiants -- the
candidate luminous blue variables (cLBVs) Sher\,25 (Brandner et
al. 1997a), HD\,168625 (Smith 2007) and MWC\,349A (Gvaramadze \&
Menten 2012), and the newly identified blue supergiant
[SBW2007]\,1 (Smith, Bally \& Walawender 2007).

There is still no complete understanding on how and in which
evolutionary phases massive stars form their circumstellar
nebulae. The nebulae can definitely originate during the late and
final (pre-supernova) phases, as evidenced by the detection of
numerous compact shells of processed material around Wolf-Rayet
stars (e.g. Chu, Treffers \& Kwitter 1983; Esteban et al. 1992;
Dopita et al. 1994; Marston 1995; Stock, Barlow \& Wesson 2011)
and the three-ring nebula associated with the SN\,1987A. Under
proper conditions, compact shells could form around red
supergiants (Mackey et al. 2014), which unlike the shells around
Wolf-Rayet stars could exist until their central stars explode as
supernovae. The majority of candidate and bona fide LBVs are also
associated with compact circular or bipolar nebulae (Nota et al.
1995; Clark, Larionov \& Arkharov 2005; Kniazev, Gvaramadze \&
Berdnikov 2015) and there is evidence that some of these stars
might be the immediate precursors of supernovae (e.g. Kotak \&
Vink 2006; Groh, Meynet \& Ekstr\"{o}m 2013; Justham,
Podsiadlowski \& Vink 2014). On the other hand, the analysis of
the heavy element abundances in circumstellar nebulae around
several LBVs by Lamers et al. (2001) led them to suggest that
these nebulae were created at the end of the main sequence.
Similarly, it was argued that enhanced mass loss at the beginning
of the blue supergiant phase was responsible for the origin of the
bipolar nebula around the cLBV Sher\,25 (Smartt et al. 2002;
Hendry et al. 2008) and a multi-shell nebula associated with the
late O-type supergiant TYC 3159-6-1 (Gvaramadze et al. 2014a).
Which of the evolutionary phases are more prone to produce
circumstellar nebulae in general and the bipolar ones in
particular still needs to be determined. Clarifying this point is
important for estimating the production rate of different types of
nebulae and for the comprehension of the mechanisms responsible
for their origin.

A large number of models have been proposed to explain the
formation of bipolar nebulae. The vast majority of them assume
that the spherically symmetric stellar wind is collimated by a
disk-like (equatorial) circumstellar environment, whose origin and
geometry could be attributed to either an asymmetric mass loss
during the preceding stellar evolutionary phase (e.g. Blondin \&
Lundqvist 1993; Martin \& Arnett 1995; Frank, Balick \& Davidson
1995; Langer, Gar\'{i}a-Segura \& Mac Low 1999) or to various
binary interaction processes (e.g. Morris 1981; Fabian \& Hansen
1979; Mastrodemos \& Morris 1999; Morris \& Podsiadlowski 2009).
In some models, the stellar wind is considered to be intrinsically
bipolar, e.g. because of nearly critical stellar rotation (Owocki
\& Gayley 1997; Maeder \& Desjacques 2001; Chita et al. 2008). It
is quite likely that several mechanisms for shaping bipolar
nebulae could be at work. The detection of new examples of bipolar
outflows from massive stars might be essential for the
identification of relevant mechanisms for their origin as well as
for a better understanding at which stellar evolutionary phase(s)
these mechanisms start to operate.

In our search for compact mid-infrared nebulae in the archival
data of the {\it Spitzer Space Telescope} (for the motivation and
some results of this search, see Gvaramadze et al. 2009, 2010a,b,
2014b; Gvaramadze, Kniazev \& Fabrika 2010c, hereafter Paper\,I),
we detected two striking examples of hourglass-like nebulae, which
were presented for the first time and named MN13 and
MN18\footnote{In the Set of Identifications, Measurements and
Bibliography for Astronomical Data (SIMBAD) data base these shells
are named [GKF2010]\,MN13 and [GKF2010]\,MN18.} in Paper\,I (see
table\,1 and figs\,3 and 10 there). MN18 and its central star (in
the following, we will use the name MN18 only for the star) are
the subject of this paper and presented in Section\,\ref{sec:neb}.
Section\,\ref{sec:obs} describes our spectroscopic and photometric
observations of MN18. The spectral classification of MN18 is given
in Section\,\ref{sec:par}, where we also determine the fundamental
parameters of this star using the radiative transfer code {\sc
cmfgen} and estimate its luminosity. Spectroscopy of the nebula is
discussed in Section\,\ref{sec:nebul}. In Section\,\ref{sec:dis},
we discuss the origin of bipolar nebulae around evolved massive
stars and compare the nebula around MN18 with other similar
bipolar nebulae. The possible parent cluster of MN18 is discussed
in Section\,\ref{sec:dis} as well.

\begin{figure*}
\begin{center}
\includegraphics[width=1.8\columnwidth,angle=0]{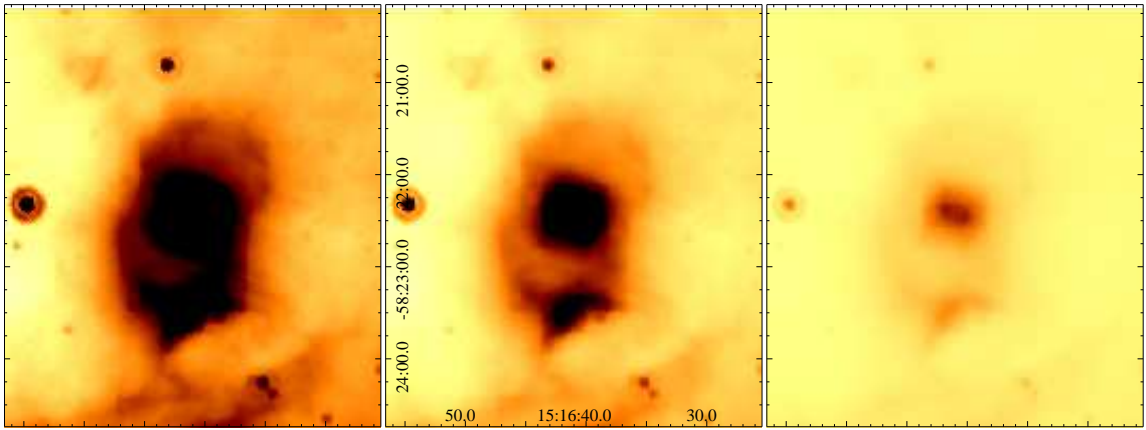}
\end{center}
\caption{{\it Spitzer} MIPS 24 $\mu$m image of the bipolar nebula
around MN18 at three intensity scales, highlighting various
details of its structure. The coordinates are in units of RA
(J2000) and Dec. (J2000) on the horizontal and vertical scales,
respectively. At an adopted distance of 5.6 kpc, 1 arcmin
corresponds to $\approx$1.6 pc.} \label{fig:neb}
\end{figure*}

\begin{figure*}
\begin{center}
\includegraphics[width=1.8\columnwidth,angle=0]{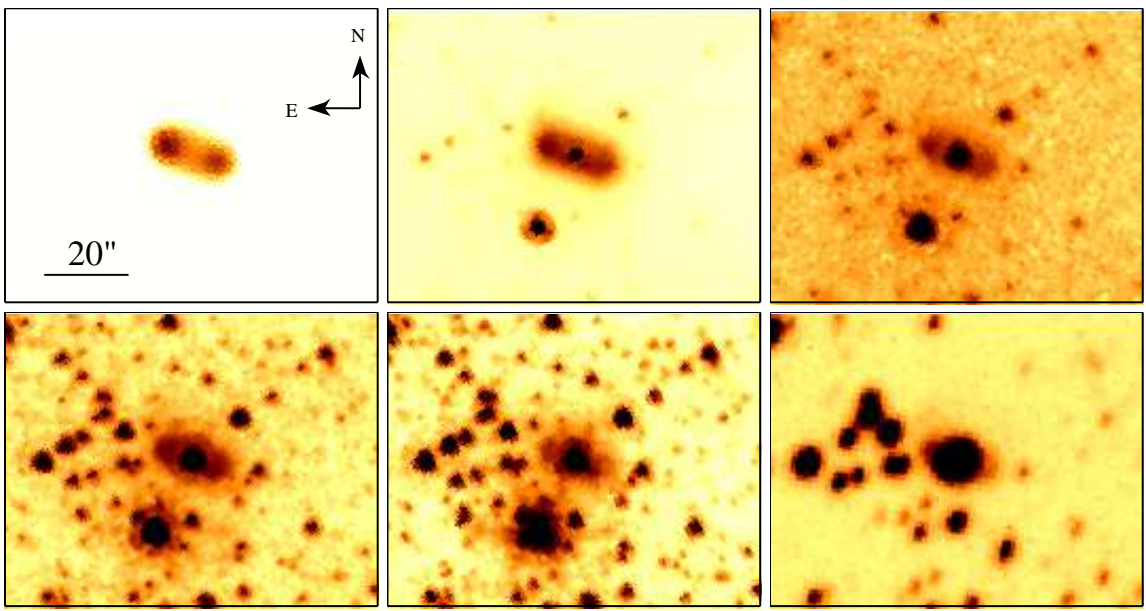}
\end{center}
\caption{From left to right, and from top to bottom: {\it Spitzer}
MIPS $24 \, \mu$m, IRAC 8, 5.8, 4.5 and 3.6\,$\mu$m, and SHS
H$\alpha$+[N\,{\sc ii}] images of the equatorial ring around MN18.
MN18 and several stars around it were proposed by Lynga (1964a) to
be members of the same star cluster (see Section\,\ref{sec:clust}
for more details).} \label{fig:neb-2}
\end{figure*}

\section{MN18 and its bipolar nebula}
\label{sec:neb}

The blue supergiant MN18 is one of many dozens of evolved massive
stars revealed through the detection of their circumstellar
nebulae with the {\it Spitzer Space Telescope} (see Table\,1 in
Paper\,I). The bipolar nebula around MN18 was discovered in
archival 24\,$\mu$m data obtained with the Multiband Imaging
Photometer for {\it Spitzer} (MIPS; Rieke et al. 2004) within the
framework of the 24 and 70 Micron Survey of the Inner Galactic
Disk with MIPS (Carey et al. 2009). This nebula appears as two
lobes protruding for about 70 arcsec in the northwest and
southeast directions from a bright waist centred on MN18. At a
distance of $d$=5.6 kpc (see Section\,\ref{sec:par}), the linear
extent of the lobes is $\approx$1.9 pc.

Fig.\,\ref{fig:neb} shows the nebula in three different intensity
scales to highlight some details of its structure. The low
intensity image of the waist of the nebula revealed two bright
spots on each side of MN18 (the star itself is not visible at
24\,$\mu$m). We interpret these spots as cross-sections of a
toroidal equatorial ring around the star visible nearly edge-on.
The northwest lobe appears to be opened along the polar axis,
while in the southeast lobe one can discern two incomplete
ellipse-like filaments oriented plane parallel to the equatorial
ring. These filaments are reminiscent of a polar ring associated
with the cLBV HD\,168625 (see fig.\,1d in Smith 2007 and fig.\,2n
in Paper\,I), but both are located on the same side of the torus.
Assuming rotational symmetry for the nebula, the aspect ratios of
the filaments suggest that the nebula is inclined with respect to
the plane of sky by $\approx$30$\degr$. Correspondingly, the
inclination angle of the stellar rotational axis to our line of
sight, $i$, is $\approx$6$0\degr$. Fig.\,\ref{fig:neb} also shows
a bright feature attached to the southwestern edge of the
southeast lobe. A possible origin of this feature is discussed in
Section\,\ref{sec:clust}.

MN18 and its equatorial ring are clearly visible in all (3.6, 4.5,
5.8 and $8.0\,\mu$m) images (see Fig.\,\ref{fig:neb-2}) obtained
with the {\it Spitzer} Infrared Array Camera (IRAC; Fazio et al.
2004) within the Galactic Legacy Infrared Mid-Plane Survey
Extraordinaire (GLIMPSE; Benjamin et al. 2003). The northeast half
of the ring is $\approx$20 per cent brighter than the southwest
one. The semimajor and semiminor axes of the nebula are
$\approx$11 and 4.5 arcsec, respectively. For $d$=5.6 kpc, the
linear radius of the waist is $\approx$0.29 pc. The 8\,$\mu$m
image also shows a ``horn" at the northeast edge of the equatorial
ring, which might correspond to the beginning of the northwest
lobe. Similar horns could also be seen in the IRAC images of the
equatorial rings of the cLBVs Sher\,25 and MN56 (see,
respectively, fig.\,2l and fig.\,4 in Paper\,I).

The IRAC images of the waist of the bipolar nebula were presented
for the first time by Kwok et al. (2008), who classified it as a
planetary nebula (PN) and named
PNG\,321.0$-$00.7\footnote{Anderson et al. (2012) used mid- and
far-infrared colours of PNG\,321.0$-$00.7 to suggest that this
object is likely a PN, while Cohen et al. (2011) confused it with
another PN candidate from Kwok et al. (2008), PNG\,306.4+00.2, and
proposed that PNG\,321.0$-$00.7 might be a reflection nebula (see
their table\,1).}. Kwok et al. (2008) also noted that
PNG\,321.0$-$00.7 has an optical counterpart in the SuperCOSMOS
H-alpha Survey (SHS; Parker et al. 2005), but did not show its
image. Fig.\,10 in Paper\,I presented the SHS H$\alpha$+[N\,{\sc
ii}] image of the waist for the first time (see also
Fig.\,\ref{fig:neb-2}).

In Paper\,I, we provisionally classified MN18 as an early B
supergiant, using a spectrum obtained with the 1.9-m telescope of
the South African Astronomical Observatory (SAAO) (see
Section\,\ref{sec:spec}). Our subsequent spectroscopy with the
Southern African Large Telescope (SALT) allowed us to refine this
classification to B1\,Ia (see Section\,\ref{sec:clas}).

The details of MN18 are summarized in Table\,\ref{tab:det}. The
$BVI_{\rm c}$ magnitudes are the mean values derived from our five
CCD measurements listed in Table\,\ref{tab:phot}. The coordinates
and the $JHK_{\rm s}$ photometry are taken from the Two-Micron All
Sky Survey (2MASS) All-Sky Catalog of Point Sources (Cutri et al.
2003). The IRAC photometry is from the GLIMPSE Source Catalog (I +
II + 3D) (Spitzer Science Center 2009).

\begin{table}
  \centering{\caption{Details of MN18.}
  \label{tab:det}
 \begin{tabular}{lr}
    \hline
  Spectral type & B1\,Ia \\
  RA(J2000) & $15^{\rm h} 16^{\rm m} 41\fs01$  \\
  Dec.(J2000) & $-$$58\degr 22\arcmin 26\farcs0$ \\
  $l$ & $321\fdg0239$ \\
  $b$ & $-$$0\fdg6992$  \\
  $B$ (mag) & 15.34$\pm$0.01 \\
  $V$ (mag) & 13.40$\pm$0.01 \\
  $I_{\rm c}$ (mag) & 10.91$\pm$0.01 \\
  $J$ (mag) & 9.22$\pm$0.02 \\
  $H$ (mag) & 8.72$\pm$0.02 \\
  $K_{\rm s}$ (mag) & 8.38$\pm$0.02 \\
  $[3.6]$ (mag) & 8.02$\pm$0.04 \\
  $[4.5]$ (mag) & 7.99$\pm$0.06 \\
  $[5.8]$ (mag) & 7.88$\pm$0.05 \\
  $[8.0]$ (mag) & 8.01$\pm$0.14 \\
  \hline
 \end{tabular}
}
\end{table}

MN18 is one of the 25 stars with $V<15$ mag, which were suggested
by Lynga (1964a) to be members of the open star cluster Ly\,3,
called Lynga\,3 in the
SIMBAD\footnote{http://simbad.u-strasbg.fr/simbad/} and
WEBDA\footnote{http://webda.physics.muni.cz/} (Mermilliod 1995)
data bases. MN18 has the sequence number 22 in table\,V of Lynga
(1964a), which gives photographic $B$ and $V$ magnitudes of 15.6
and 13.3, respectively, for this star. The relationship between
MN18 and Lynga\,3 is further discussed in
Section\,\ref{sec:clust}.

\section{MN18: observations}
\label{sec:obs}

\subsection{Spectroscopic observations}
\label{sec:spec}

\begin{table*}
  \caption{Journal of the observations.}
  \label{tab:log}
  \renewcommand{\footnoterule}{}
  \begin{tabular}{llccccccccc}
      \hline
      Spectrograph & Date & Exposure & Spectral scale & Spatial scale & Slit & Spectral range \\
      & & (min) & (\AA \, pixel$^{-1}$) & (arcsec pixel$^{-1}$) & (arcsec) & (\AA) \\
      \hline
      Cassegrain (SAAO 1.9-m) & 2009 May 27$^{a}$ & 3$\times$15 & 2.3 & 1.4 & 1.8$\times$180 & 4200$-$8100 \\
      RSS (SALT) & 2012 March 2 & 5 & 0.97 & 0.25 & 1.5$\times$480 & 4200$-$7300 \\
      RSS (SALT) & 2013 April 26 & 15 & 0.27 & 0.51 & 1.25$\times$480 & 6035$-$6900 \\
      RSS (SALT) & 2013 April 27 & 15 & 0.27 & 0.51 & 1.25$\times$480  & 3500$-$9500 \\
      FEROS (MPG 2.2-m) & 2014 March 8 & 3$\times$30 & 0.038 & -- & -- & 3815$-$5440 \\
      FEROS (MPG 2.2-m) & 2014 March 9 & 6$\times$30 & 0.038 & -- & -- & 3815$-$5440 \\
      \hline
      \MC{4}{l}{{\it Note.} $^{a}$Presented for the first time in Paper\,I.}
    \end{tabular}
    \end{table*}

The first spectrum of MN18 was obtained with the SAAO 1.9-m
telescope on 2009 May 27. The observations were performed with the
Cassegrain spectrograph using a slit of 3 arcmin $\times$ 1.8
arcsec and grating with 300 lines mm$^{-1}$. The slit was oriented
at a position angle (PA) of PA=80$\degr$, measured from north to
east. This spectral setup covered a wavelength range of
$\approx$4200--8100\,\AA\ with a reciprocal dispersion of
$\approx$2.3\,\AA\,pixel$^{-1}$ and a spectral resolution full
width at half-maximum (FWHM) of $\approx$7\,\AA. Three exposures
of 900\,s were taken with a seeing of $\approx$2 arcsec.

The obtained spectrum (see fig.\,9 of Paper\,I) revealed numerous
H and He\,{\sc i} lines in absorption and prominent nebular
emissions -- the H$\alpha$ line flanked by the [N\,{\sc ii}]
$\lambda\lambda$6548, 6584 lines. Unfortunately, the stellar and
the nebular spectra were too weak to allow us to obtain a reliable
spectral classification of MN18 and to derive any useful
information on the nebula.

To classify MN18 and to study the nebula, we obtained a new
spectrum (we call it low-resolution hereafter) with the SALT
(Buckley, Swart \& Meiring 2006; O'Donoghue et al. 2006) on 2012
March 2, using the Robert Stobie Spectrograph (RSS; Burgh et al.
2003; Kobulnicky et al. 2003) in the long-slit mode with a slit of
8 arcmin$\times$1.5 arcsec. The slit was oriented along the major
axis of the optical nebula, i.e. at PA=72$\degr$. The PG900
grating was used to cover the spectral range of 4200$-$7300~\AA \,
with a final reciprocal dispersion of $0.97$\,\AA\,pixel$^{-1}$
(FWHM$\approx$5.50\,\AA). A 300\,s spectrum was taken with
2$\times$2 binning. The RSS uses a mosaic of three
2048$\times$4096 CCDs and the final spatial scale for this
observation was $0\farcs25$\,pix$^{-1}$. The seeing during the
observation was $\approx$1.3 arcsec. A Xe lamp arc spectrum was
taken immediately after the science frame. A spectrophotometric
standard star was observed during twilight time for relative flux
calibration.

To study the velocity distribution along the optical nebula, two
additional higher resolution long-slit spectra (high-resolution
spectra hereafter) were obtained with the RSS using the grating
PG2300 on 2013 April 26 and 27. A slit with a width of 1.25 arcsec
was used in both observations with the same PA as earlier. A
spectral range of 6035$-$6900\,\AA\ was covered with a final
reciprocal dispersion of $0.27$\,\AA\,pixel$^{-1}$ and a FWHM
spectral resolution of 1.33\,\AA. One 900\,s spectrum was taken
with a 2$\times$4 binning each night with a spatial scale of
$0\farcs51$\,pix$^{-1}$ along the slit. The seeing during the
observations was 1.5 arcsec. A Ne lamp arc spectrum and a set of
Quartz Tungsten Halogen flats were taken immediately after the
science frames, and a spectrophotometric standard star was
observed during twilight time for relative flux calibration.

The primary reduction of the SALT data was carried out with the
SALT science pipeline (Crawford et al. 2010). After that, the bias
and gain corrected and mosaiced long-slit data were reduced in the
way described in Kniazev et al. (2008). Absolute flux calibration
is not feasible with SALT because the unfilled entrance pupil of
the telescope moves during the observation. However, a relative
flux correction to recover the spectral shape was done using the
observed spectrophotometric standard.

For spectral modelling and to derive the rotational velocity, we
obtained high-resolution spectra of MN18 with the 2.2-m Max Planck
Gesellschaft telescope at La Silla (European Southern Observatory;
ESO). The data were taken on 2014 March 8 and 9 with the Fibre-fed
Extended Range Optical Spectrograph (FEROS; Kaufer et al. 1999) --
a high-resolution \'{E}chelle spectrograph with an aperture of 2
arcsec, which provides a resolution of 0.038\,\AA\,pixel$^{-1}$. A
2048$\times$4096 CCD was used to record 39 orders covering the
spectral range of 3500--9500\,\AA. Nine 1800\,s exposures were
obtained, of which three were taken during the first night. The
seeing during the observation was below 1.5 arcsec in both nights.
Wavelength calibration frames were obtained with ThArNe lamps. A
spectrophotometric standard star was observed directly before the
target observation in order to perform a flux calibration.

The reduction of the FEROS data, including flatfield correction,
bias and background subtraction, order tracing, wavelength and
flux calibration, was performed manually with the FEROS Data
Reduction Software based on ESO-MIDAS.

The log of our spectroscopic observations is listed in
Table\,\ref{tab:log}.

\subsection{Photometry}
\label{sec:phot}

To detect possible photometric variability of MN18, we determined
its $B, V$ and $I_{\rm c}$ magnitudes on CCD frames obtained with
the 76-cm telescope of the SAAO during our five observing runs in
2009--2014. We used an SBIG ST-10XME CCD camera equipped with
$BVI_{\rm c}$ filters of the Kron-Cousins system (see e.g.
Berdnikov et al. 2012). We also calibrated the 1.9-m telescope
spectrum and synthesized its $V$ magnitude in the way described in
Kniazev et al. (2005). Additionally, the $V$ magnitude of MN18 was
measured on $V$ band images taken before each SALT spectroscopic
observation. For this we used secondary photometric standards
calibrated with the 76-cm telescope data. Using the same data, we
also re-calibrated the photographic $B$ and $V$ magnitudes of MN18
from Lynga (1964a). The resulting photometry is presented in
Table\,\ref{tab:phot}.

From Table\,\ref{tab:phot} it follows that MN18 did not experience
major changes in its brightness: during the last five years its
$B$, $V$ and $I_{\rm c}$ magnitudes changed by $\la$0.1 mag, while
its $B-V$ and $B-I_{\rm c}$ colours remained almost constant with
mean values of 1.94$\pm$0.01 and 4.43$\pm$0.01 mag, respectively.

\begin{table}
  \caption{Photometry of MN\,18.}
  \label{tab:phot}
  \renewcommand{\footnoterule}{}
  \begin{center}
  \begin{minipage}{\textwidth}
    \begin{tabular}{lccc}
      \hline
      Date & $B$ & $V$ & $I_{\rm c}$ \\
      \hline
      1963 April 4--17$^{(a)}$ & 15.46$\pm$0.25 & 13.28$\pm$0.21 & --             \\
      2009 May 27$^{(b)}$      & --             & 13.35$\pm$0.03 & --             \\
      2010 January 22$^{(c)}$  & 15.36$\pm$0.01 & 13.42$\pm$0.01 & 10.94$\pm$0.01 \\
      2012 March 2$^{(d)}$     & --             & 13.33$\pm$0.04 & --             \\
      2012 May 6$^{(c)}$       & 15.30$\pm$0.01 & 13.36$\pm$0.01 & 10.87$\pm$0.01 \\
      2013 January 13$^{(c)}$  & 15.34$\pm$0.01 & 13.41$\pm$0.01 & 10.90$\pm$0.01 \\
      2013 April 26$^{(d)}$    & --             & 13.44$\pm$0.04 & --             \\
      2014 January 18$^{(c)}$  & 15.33$\pm$0.02 & 13.43$\pm$0.01 & 10.92$\pm$0.01 \\
      2014 April 8$^{(c)}$     & 15.38$\pm$0.02 & 13.41$\pm$0.02 & 10.94$\pm$0.02 \\
      \hline
    \end{tabular}
    \end{minipage}
    \end{center}
    {\it Note:} $^a$Lynga (1964a); $^b$1.9-m telescope; $^c$76-cm telescope; $^d$SALT.
\end{table}

\section{MN18: spectral analysis and stellar parameters}
\label{sec:par}

\subsection{Classification of MN18 and changes in
the H$\alpha$ line}
\label{sec:clas}

\begin{figure*}
\begin{center}
\includegraphics[width=11cm,angle=270,clip=]{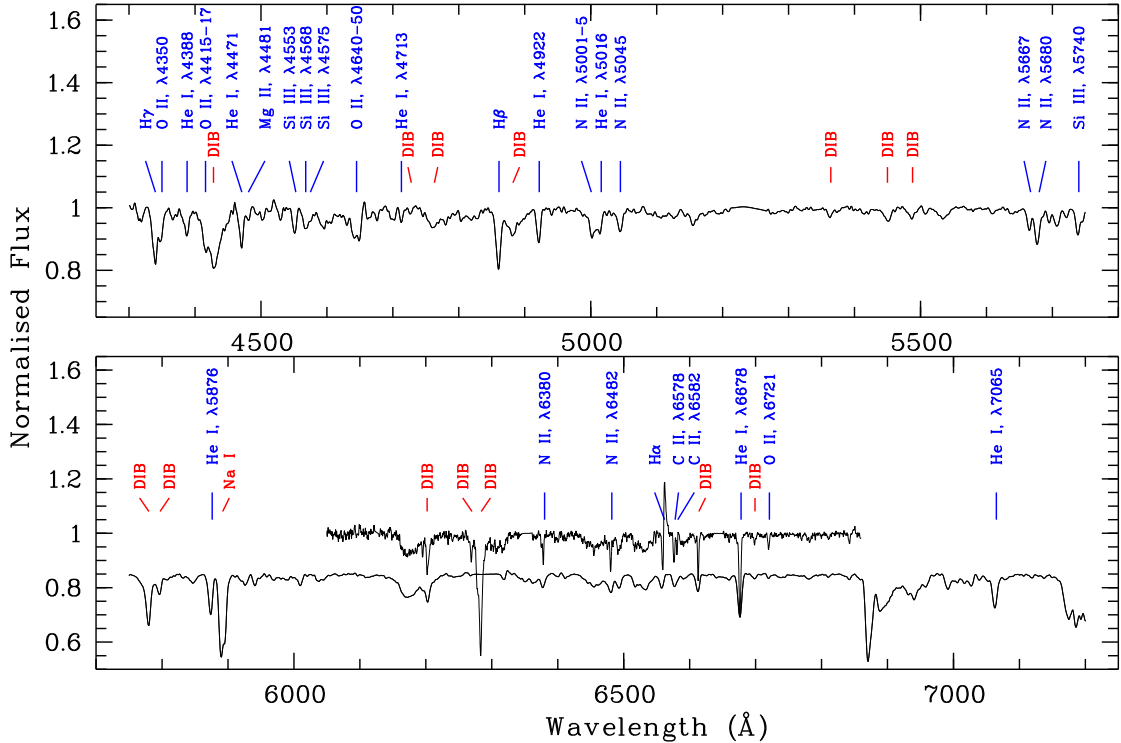}
\end{center}
\caption{Normalized low-resolution spectrum of MN18 obtained with
the SALT on 2012 March 2 with the principal lines and most
prominent DIBs indicated. The lower panel also shows the
high-resolution spectrum obtained with the SALT on 2013 April 27.}
\label{fig:specl}
\end{figure*}

The normalized low- and high-resolution SALT spectra of MN18 are
presented in Fig.\,\ref{fig:specl}, with the principal lines and
most prominent diffuse interstellar bands (DIBs) indicated. All
wavelengths are given in air. Equivalent widths (EWs), FWHMs and
heliocentric radial velocities (RVs) of the main lines in the
low-resolution spectrum (measured applying the {\sc midas}
programs; see Kniazev et al. 2004 for details) are summarized in
Table\,\ref{tab:int}.

\begin{table}
\caption{EWs, FWHMs and RVs of main lines in the low-resolution
spectrum of MN18.} \label{tab:int}
\renewcommand{\footnoterule}{}
  \begin{center}
  \begin{minipage}{\textwidth}
\begin{tabular}{lccc}
\hline
$\lambda_{0}$(\AA) Ion & EW($\lambda$) & FWHM($\lambda$) & RV \\
& (\AA) & (\AA) & ($\kms$) \\
\hline
4340\ H$\gamma$\        &  1.61$\pm$0.04 &  8.68$\pm$0.41 &  -55$\pm$9 \\
4388\ He\ {\sc i}\      &  0.26$\pm$0.02 &  8.46$\pm$0.67 &  -22$\pm$8 \\
4471\ He\ {\sc i}\      &  0.94$\pm$0.04 &  7.03$\pm$0.36 &  -51$\pm$8 \\
4481\ Mg\ {\sc ii}\     &  0.18$\pm$0.02 &  5.33$\pm$0.90 &    2$\pm$8 \\
4553\ Si\ {\sc iii}\    &  0.56$\pm$0.02 &  6.34$\pm$0.23 &  -63$\pm$8 \\
4713\ He\ {\sc i}\      &  0.18$\pm$0.01 &  4.85$\pm$0.21 &  -11$\pm$8 \\
4861\ H$\beta$\         &  1.61$\pm$0.05 &  9.19$\pm$0.31 &  -48$\pm$8 \\
4922\ He\ {\sc i}\      &  0.59$\pm$0.01 &  6.40$\pm$0.17 &  -37$\pm$7 \\
5045\ N\ {\sc ii}\      &  0.31$\pm$0.01 &  8.85$\pm$0.39 &  -44$\pm$7 \\
5667\ N\ {\sc ii}\      &  0.53$\pm$0.02 &  8.13$\pm$0.18 &  -68$\pm$6 \\
5680\ N\ {\sc ii}\      &  0.82$\pm$0.02 &  8.13$\pm$0.18 &  -104$\pm$6 \\
5740\ Si\ {\sc iii}\    &  0.67$\pm$0.01 &  7.30$\pm$0.17 &  -34$\pm$6 \\
5876\ He\ {\sc i}\      &  0.90$\pm$0.02 &  6.91$\pm$0.17 &  -78$\pm$6 \\
6563\ H$\alpha$$^{(a)}$\ & 0.20$\pm$0.01 & 6.70$\pm$0.21 & $-$221$\pm$5 \\
6678\ He\ {\sc i}\      &  1.15$\pm$0.02 &  6.51$\pm$0.15 &  -69$\pm$6 \\
7065\ He\ {\sc i}\      &  1.13$\pm$0.04 &  9.41$\pm$0.41 &  -90$\pm$5 \\
\hline
\end{tabular}
    \end{minipage}
    \end{center}
    {\it Note:} $^a$The RV of this line is noticeably affected by the P\,Cygni absorption.
\end{table}

Fig.\,\ref{fig:specl} shows that the spectrum of MN18 is dominated
by absorption lines of H and He\,{\sc i}. In the high-resolution
spectrum the H$\alpha$ line is in emission and shows a P\,Cygni
profile. Other prominent absorption lines are those of N\,{\sc ii}
$\lambda\lambda$5001-5, 5045, 5667, 5680, 6380, 6482, O\,{\sc ii}
$\lambda\lambda$4350, 4415-17, 4640-50, 6721, Mg\,{\sc i}
$\lambda$4471, Si\,{\sc iii} $\lambda\lambda$4553-68-75, 4650
5740, and C\,{\sc ii} $\lambda\lambda$6578, 6583. No He\,{\sc ii}
lines are present in the spectrum, which implies that MN18 is of
spectral type of B1 or later (Walborn \& Fitzpatrick 1990).

Using the spectral atlas of OB stars by Walborn \& Fitzpatrick
(1990), we classify MN18 as a B1\,Ia star. The Ia luminosity class
also follows from the measured EW(H$\gamma$) of 1.61$\pm$0.04 \AA
\, and the calibrations of Balona \& Crampton (1974; see their
table\,V).

The high interstellar extinction towards MN18 (see
Sections\,\ref{sec:mod} and \ref{sec:neb-phy}) is manifested in
numerous DIBs, the most prominent of which are at 4429, 4762,
4881, 5780, 5797, 6203, 6269, 6283 and 6614 \AA.

\begin{figure}
\begin{center}
\includegraphics[angle=270,width=1.0\columnwidth,clip=]{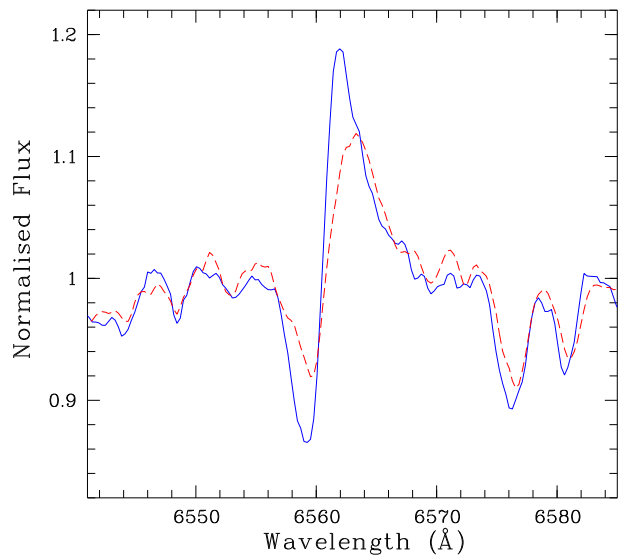}
\end{center}
\caption{Daily time-scale line profile variability in H$\alpha$
and the C\,{\sc ii} $\lambda\lambda$6578, 6583 doublet: 2013 April
26 (solid line) and 2013 April 27 (dashed line).} \label{fig:Ha}
\end{figure}

Inspection of the spectra of MN18 obtained in 2009--2014 did not
reveal significant changes in their appearance, except for the
variability of the H$\alpha$ line. It was in emission in 2009 and
2013, in absorption in 2012, and in transition between these two
states in 2014, with the line centre in absorption and the line
wings in emission. Comparison of the SALT high-resolution spectra
obtained during two consecutive nights in 2013 shows that the
H$\alpha$ line (and the C\,{\sc ii} $\lambda\lambda$6578, 6583
doublet as well) changes its shape on a daily time-scale (see
Fig.\,\ref{fig:Ha}), which is typical of early-type supergiants
(e.g. Morel et al. 2004; Martins et al. 2015). This variability
introduces an ambiguity in mass-loss estimates based on
synthesizing the H$\alpha$ line with model stellar atmosphere
codes (see Section\,\ref{sec:mod}).

Using the 1.9-m telescope and SALT spectra, we searched for shifts
in RV as a result of possible binarity of MN18 and found that all
measurements agree with each other within their error margins.

\subsection{Spectral modelling}
\label{sec:mod}

\begin{figure*}
\begin{center}
\includegraphics[width=16cm,angle=0]{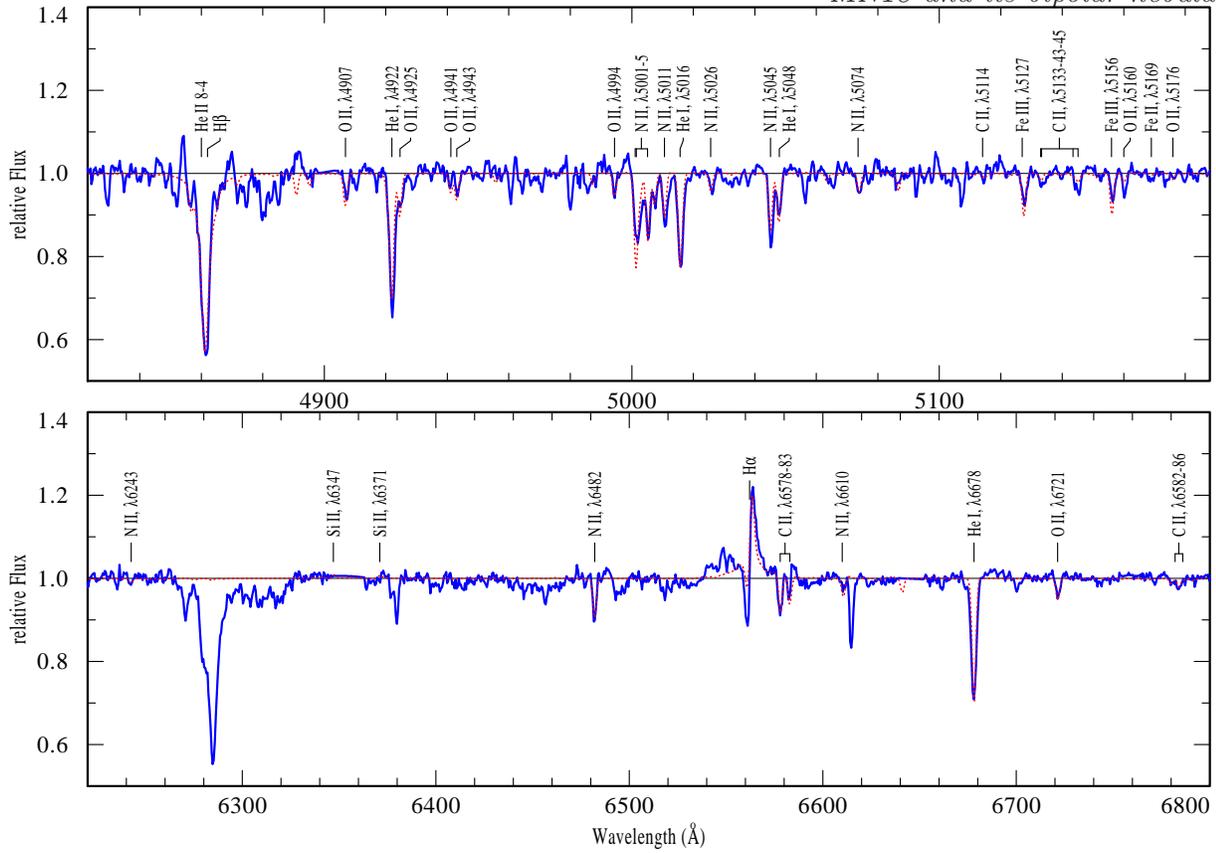}
\end{center}
\caption{Parts of the normalized FEROS (upper panel) and high
resolution SALT (bottom panel) spectra of MN18 (shown by blue
solid lines), compared with the best-fitting model spectrum (red
dotted line) with the parameters as given in
Table\,\ref{tab:par}.} \label{fig:fit}
\end{figure*}

\begin{figure*}
\begin{center}
\includegraphics[width=16cm,angle=0]{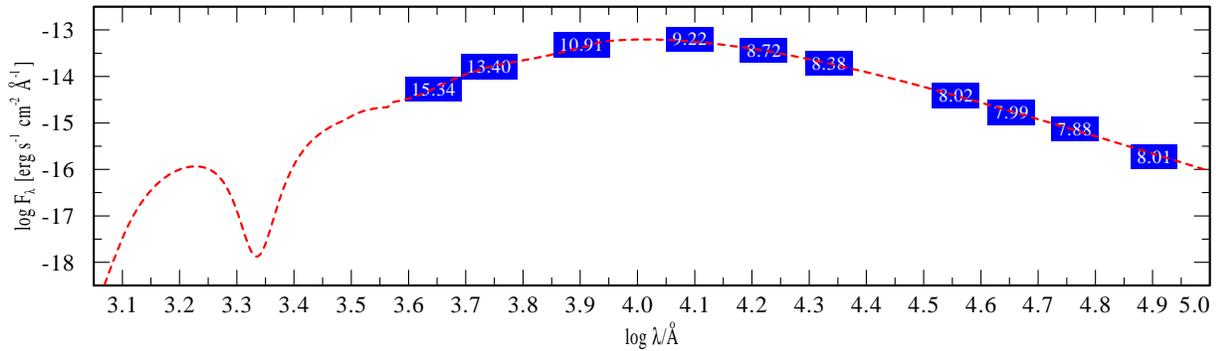}
\end{center}
\caption{Observed flux distribution of MN18 in absolute units,
based on the photometric measurements (blue rectangles) compiled
in Table\,\ref{tab:det}, compared to the emergent flux of the
reddened model spectrum (red dashed line) with the parameters as
given in Table\,\ref{tab:par}.} \label{fig:sed}
\end{figure*}

For the spectral analysis we used spectra from three epochs: the
low- and high-resolution SALT spectra taken on 2012 March 2 and
2013 April 27, respectively, and the echelle FEROS spectrum taken
on 2014 March 9. Our analysis is based on synthetic spectra
computed with the non-LTE stellar atmosphere code {\sc cmfgen}
(Hillier \& Miller 1998). The following model atoms and ions were
taken into account: H\,{\sc i}, He\,{\sc i-ii}, C\,{\sc i-iv},
N\,{\sc iii-v}, O\,{\sc i-v}, Mg\,{\sc ii}, Al\,{\sc ii}, Si\,{\sc
ii-iv} and Fe\,{\sc i-v}. Because we are mainly interested in the
evolutionary status of MN18, we varied only the abundances of He,
C, N and O, while the abundances of other metals are set at solar
values (taken from Asplund et al. 2005). Since our spectra did not
cover the UV resonance lines, we cannot resolve the degeneracy
between the wind velocity law, $\beta$, and a wind volume filling
factor of optically thin (micro) clumps, $f_{\rm v}$. Similar line
profiles for H$\alpha$ can be achieved by varying $\beta$ and/or
$f_{\rm v}$. Based on our model computations, $\beta$ ranges
between 1.5 and 2.5, and $f_{\rm v}$ between 0.25 and 1.0. In
order to estimate the error bars in the effective temperature,
$T_{\rm eff}$, mass-loss rate, $\dot{M}$, and surface gravity,
$\log g$, we computed a grid of models around the preferred
values. The best-fitting model is shown in Fig.\,\ref{fig:fit}.

$T_{\rm eff}$ is based on the ratio of the Fe\,{\sc ii} and {\sc
iii} lines and the intensity of the He\,{\sc i}, N\,{\sc ii} and
Si\,{\sc ii} ones, and is 21.1$\pm$1.0\,kK. We employed H$\beta$
to determine $\log g$=2.75$\pm$0.20. The uncertainty of $\log g$
is rather large as a result of low signal to noise ratio in the
FEROS and low resolution in the SALT spectra. The terminal
velocity, $v_\infty$, is difficult to constrain due to the weak
emission of H$\alpha$. No other suitable lines are available in
the covered wavelength range. Therefore, the uncertainty is large
and we estimate $v_\infty$ of $\approx$1400$\pm$$500 \, \kms$. For
comparison, we used the relation between $v_\infty$ and the
photospheric escape velocity, $v_{\rm esc}$, from Lamers, Snow \&
Lindholm (1995), which for $T_{\rm eff}=21.1$\,kK, $\log g$=2.75
and the bolometric luminosity of $\log(L/\lsun)$=5.42 (see below)
gives $v_\infty/v_{\rm esc}=2.6$ and $v_\infty$=1370$\pm$$250 \,
\kms$.

The mass-loss rates for the three epochs were derived from
H$\alpha$. Over these epochs the profile of H$\alpha$ changed and,
correspondingly, $\dot{M}$ as well. The obtained values for the
individual epochs with $\log(L/\lsun)$=5.42, $v_\infty$=$1400 \,
\kms$, $\beta$=2.0 and $f_{\rm v}$=1.0 are
$\log(\dot{M}/\myr)$=$-$6.38$\pm$0.25\footnote{This is an upper
limit because the H$\alpha$ line was in absorption in 2012.} (2012
March 2), $-$6.17$\pm$0.25 (2013 April 27), and $-$6.37$\pm$0.25
(2014 March 9). We observe a variation of $\dot{M}$ by a factor of
$\approx$1.6. The errors in $\dot{M}$ account for the
uncertainties in $\log(L/\lsun)$, $v_{\infty}$, $\beta$-parameter
and micro-clumping. Petrov et al. (2014) found that the optical
depth of the H$\alpha$ line below the bi-stability jump ($T_{\rm
eff}$$\approx$22.5\,kK) increases with decreasing $T_{\rm eff}$.
Correspondingly, the profile of this line becomes sensitive to
macro-clumping and the derived $\dot{M}$ could be underestimated.
However, the effective temperature of MN18 is close to that of the
bi-stability jump, while the $v_{\infty}/v_{\rm esc}$ ratio of 2.6
suggests that MN18 has wind properties expected for a star on the
hot side of this jump. Therefore, it appears likely that the
influence of the macro-clumping on the mass loss rate of MN18 is
small, i.e., negligible in comparison to the overall
uncertainties.

The CNO abundances are based on the ratio of the C, N and O
spectral lines. Table\,\ref{tab:par} shows that C is reduced by a
factor of $\approx$3.5 and N is enriched by $\approx$4 times due
to the CNO cycle. The abundance of O is still close to its initial
value within the error margins. At the given wavelength range the
H and He lines are insensitive to changes in the He abundance, so
that the He mass fraction could not be derived.

To estimate the reddening towards MN18, we match the model
spectral energy distribution (SED) with the observed one in the
$B$, $V$ and $K_{\rm s}$ bands (see Table\,1). We extracted
intrinsic $B-V$ and $V-K_{\rm s}$ colours from the model flux by
applying approximate filter functions\footnote{The filter
functions were approximated for each filter system individually
using Gaussian and boxcar functions with appropriate widths and
wavelengths.} and match the resulting values of the colour
excesses $E(B-V)$ and $E(V-K_{\rm s})$ as described by
Bestenlehner et al. (2011) to determine the total-to-selective
absorption ratio $R_{\rm V}$ and $V$ band extinction $A_{\rm V}$.
We derive $\log(L/\lsun)$=5.42$\pm$0.2 adopting the absolute
visual magnitude of $M_{\rm V}$=$-6.8$$\pm$0.5 typical of B\,Ia
supergiants (Crowther et al. 2006). Accounting for the uncertainty
in $T_{\rm eff}$ and photometric errors, the uncertainty in
$\log(L/\lsun)$ sums up to $\pm$0.3 dex. Even though we find a
variability in H$\alpha$, the temperature and bolometric
correction (BC=$-$2.01 mag\footnote{Note that almost the same
value of BC ($-$2.03 mag) follows from relation between BC and
$T_{\rm eff}$ of B supergiants given in Crowther et al. (2006).})
are nearly unchanged, so that we can assume no significant
variations in the luminosity over time. The resulting SED well
matches the photometry from optical to mid-infrared wavelengths,
and no infrared excess is observed (see Fig.\,\ref{fig:sed}).

By converting the luminosity range into distances, we found that
MN18 lies at a distance of $d$$\approx$5.6$^{+1.5} _{-1.2}$ kpc.
The line of sight towards MN18 ($l$$\approx$321$\degr$) first
crosses the Carina-Sagittarius Arm at $\approx$0.7--1.5 kpc and
then the Crux-Scutum one at $\approx$3.6--4.4 kpc (Vall\'{e}e
2014). Our distance estimate implies that MN18 is located either
in the Crux-Scutum Arm or in the interarm region. In
Section\,\ref{sec:clust}, we argue that the former possibility is
more plausible. Note that the luminosity of MN18 and its $R_*$ and
$\dot{M}$ (given in Table\,\ref{tab:par}) can be scaled to
different distances as $\propto$$d^2$, $\propto$$d$ and
$\propto$$d^{3/2}$, respectively (Schmutz, Hamann \& Wessolowski
1989).

\begin{table}
\caption{Stellar and extinction parameters for MN18. }
\label{tab:par}
\renewcommand{\footnoterule}{}
  \begin{center}
  \begin{minipage}{\textwidth}
    \begin{tabular}{lr}
    \hline
$T_{\rm eff}$ (kK)              & 21.1$\pm$1.0 \\
$\log(L/\lsun)$                 & 5.42$\pm$0.30 \\
$R_*$ ($\rsun$)                 & 38.5$^{+18.5} _{-12.7}$ \\
$\log g$                        & $2.75\pm$0.20 \\
$\log(\dot{M}/\myr)^{(a)}$      & $-$6.38$\pm$0.25 \\
$\log(\dot{M}/\myr)^{(b)}$      & $-$6.17$\pm$0.25 \\
$\log(\dot{M}/\myr)^{(c)}$      & $-$6.37$\pm$0.25 \\
$\beta$                         & 1.5 to 2.5 \\
$f_{\rm v}$                     & 0.25 to 1.0 \\
$v_\infty$ ($\kms$)             & 1400$\pm$500 \\
$v_{\rm eq}$ ($\kms$)           & 104$\pm$7 \\
C (mass fraction)               & 6.5$\pm$2.5$\times10^{-4}$ \\ 
N (mass fraction)               & 2.4$^{+2.1} _{-1.3}\times10^{-3}$ \\ 
O (mass fraction)               & $5.4^{+1}_{-2}\times10^{-3}$ \\ 
$E(B-V)$ (mag)                  & 1.97$\pm$0.10 \\
$R_{\rm V}$                     & 3.26$\pm$0.10 \\
$M_{\rm V}$ (mag) (adopted)     & $-$6.8$\pm$0.5 \\
$d$ (kpc)                       & $5.6^{+1.5}_{-1.2}$ \\
\hline
\end{tabular}
\end{minipage}
\end{center}
{\it Note:} $^a$2012 March 2; $^b$2013 April 27; $^c$2014 March 9.
\end{table}

Using the FEROS spectra, we estimated the projected rotational
velocity, $v\sin i$, of MN18. For this, we used the relations
between the FWHMs of the He\,{\sc i} $\lambda\lambda$4387, 4471
lines and $v\sin i$ given in Steele, Negueruela \& Clark (1999).
For FWHM(4388)=2.24$\pm$0.25\,\AA \, and
FWHM(4471)=2.10$\pm$0.13\,\AA, we found $v\sin
i$$\approx$94.1$\pm$$10.5 \, \kms$ and 86.6$\pm$$5.4 \, \kms$,
respectively, with a mean value of 90.4$\pm$$5.9 \, \kms$, which
for $i$=$60\degr$ (see Section\,\ref{sec:neb}) translates into the
equatorial rotational velocity of $v_{\rm eq}$=104$\pm$$7 \,
\kms$. These estimates should be considered as an upper limit to
the rotational velocity because the macroturbulence could
contribute to the line broadening.

A summary of the stellar and extinction parameters is given in
Table\,\ref{tab:par}.

\section{Spectroscopy of the equatorial ring}
\label{sec:nebul}

\subsection{Physical conditions and elemental abundances}
\label{sec:neb-phy}

\begin{figure}
\begin{center}
\includegraphics[angle=0,width=1.0\columnwidth,clip=]{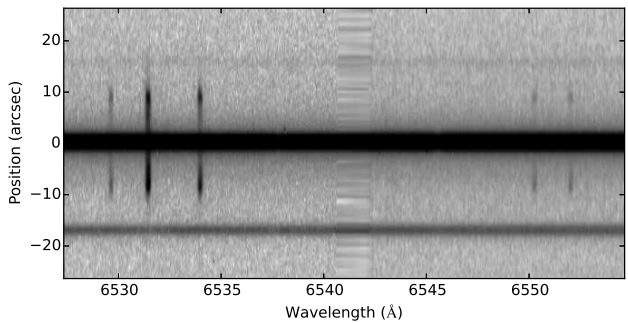}
\end{center}
\caption{A part of the 2D high-resolution spectrum of the
equatorial ring around MN18, showing nebular emission lines (from
left to right) of [N\,{\sc ii}] $\lambda$6548, H$\alpha$, [N\,{\sc
ii}] $\lambda$6548 and [S\,{\sc ii}] $\lambda\lambda$6717, 6731.}
\label{fig:2D}.
\end{figure}

As noted in Section~\ref{sec:spec}, all SALT spectra were obtained
with the slit oriented along the major axis of the equatorial
nebula. In two-dimensional (2D) spectra the nebula appears as
emission lines of H$\beta$, H$\alpha$, [\ion{N}{ii}]
$\lambda\lambda$6548, 6584 and [\ion{S}{ii}] $\lambda\lambda$6717,
6731, visible on both sides of MN18. Note that in the region of
the H$\alpha$ line the nebula is more than 5 mag fainter than the
star and therefore it does not affect the stellar photometry
appreciably. A part of one of the two 2D high-resolution spectra
is shown in Fig.\,\ref{fig:2D}. Besides the nebular spectrum and
the spectrum of MN18, Fig.\,\ref{fig:2D} also shows spectra of two
other stars on the RSS slit, which are located at almost equal
distances on both sides of MN18. In Section\,\ref{sec:clust} we
show that the brighter star is of late O-/early B-type.

\begin{figure}
\begin{center}
\includegraphics[angle=0,width=1.0\columnwidth,clip=]{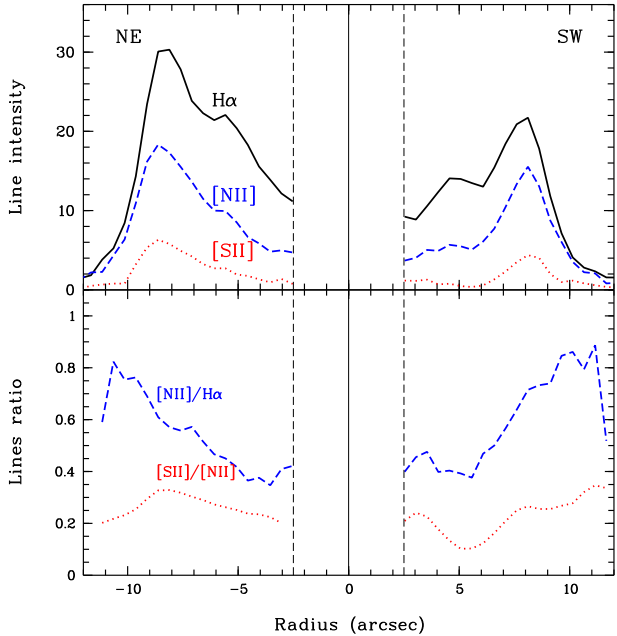}
\end{center}
\caption{Variation of emission line relative fluxes and their
ratios along the slit. The solid vertical line corresponds to the
position of MN18, while vertical dashed lines at $\pm2.5\arcsec$
from the solid line mark the area where no nebular emission was
detected. NE$-$SW direction of the slit is shown. {\it Upper
panel:} The relative fluxes of the H$\alpha$, [\ion{N}{ii}]
$\lambda$6584 and [\ion{S}{ii}] $\lambda$6717+$\lambda$6731 lines
after continuum subtraction. {\it Bottom panel:} [\ion{N}{ii}]
$\lambda$6584/H$\alpha$ and
[\ion{S}{ii}]\,$\lambda$6717+$\lambda$6731/[\ion{N}{ii}]
$\lambda$6584 line ratios.} \label{fig:Lines}
\end{figure}

The upper and bottom panels of Fig.\,\ref{fig:Lines} plot the
nebular emission line intensities and their ratios along the slit.
From the upper panel it follows that the nebula can be traced up
to $\pm$11 arcsec (at a ten per cent level of the H$\alpha$ peak
intensity) from MN18, which corresponds with the extent of the
equatorial ring in the {\it Spitzer} and SHS images (see
Fig.\,\ref{fig:neb-2}). This panel also shows that the line
intensities peak at $\approx$8.5 arcsec from the star, which
reflects the toroidal topology of the nebula.

The lower panel of Fig.\,\ref{fig:Lines} shows that the
[\ion{N}{ii}] $\lambda$6584/H$\alpha$ line ratio increases with
distance from MN18 and peaks at the edge of the nebula. Using the
prescription by Dopita (1973), one finds that
[\ion{N}{ii}]/H$\alpha$ is proportional to $n_{\rm e} ^{-1} T_{\rm
e} ^{1/2} \exp(-21\,855/T_{\rm e})$, where $n_{\rm e}$ and $T_{\rm
e}$ are the electron number density and temperature, respectively.
Thus, the behaviour of [\ion{N}{ii}] $\lambda$6584/H$\alpha$ could
be explained (at least in part) by the temperature increase of the
photoionized gas towards the outer edge of the nebula (caused by
the hardening of the ionizing radiation field), provided that the
density is constant throughout the nebula.

\begin{table}
\centering{\caption{Line intensities of the MN\,18 nebula.}
\label{tab:int2}
\begin{tabular}{lrc} \hline
\rule{0pt}{10pt}
$\lambda_{0}$(\AA) Ion    & F($\lambda$)/F(H$\beta$)&I($\lambda$)/I(H$\beta$) \\
\hline
4861\ H$\beta$\         &  1.000$\pm$0.236 & 1.000$\pm$0.249 \\
6548\ [N\ {\sc ii}]\    &  7.083$\pm$1.183 & 0.578$\pm$0.111 \\
6563\ H$\alpha$\        & 36.750$\pm$6.136 & 2.947$\pm$0.564 \\
6584\ [N\ {\sc ii}]\    & 21.792$\pm$3.647 & 1.705$\pm$0.327 \\
6717\ [S\ {\sc ii}]\    &  3.708$\pm$0.631 & 0.248$\pm$0.049 \\
6731\ [S\ {\sc ii}]\    &  3.750$\pm$0.637 & 0.247$\pm$0.049 \\
  & & \\
$C$(H$\beta$)           & \MC {2}{c}{3.30$\pm$0.22} \\
$E(B$$-$$V$)           & \MC {2}{c}{2.24$\pm$0.15 mag} \\
  & & \\
$n_{\rm e}$([S\,{\sc ii }]) & \MC {2}{c}{560$^{+450}_{-270}$ cm$^{-3}$} \\
\hline
\end{tabular}
 }
\end{table}

\begin{figure}
\begin{center}
\includegraphics[angle=270,width=1.0\columnwidth,clip=]{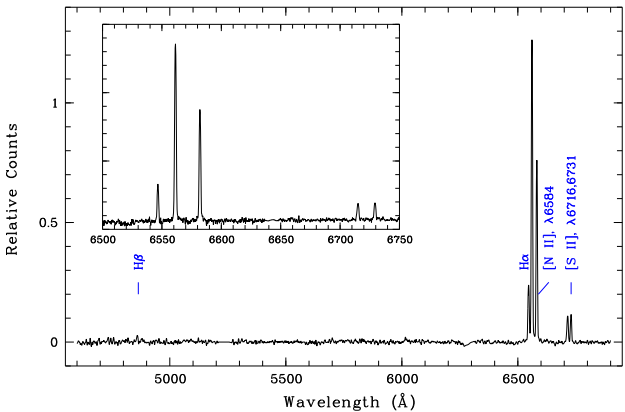}
\end{center}
\caption{1D low-resolution spectrum of the equatorial nebula
around MN18. The insert shows a part of the high-resolution
spectrum of the nebula taken on 2013 April 27.} \label{fig:Neb}
\end{figure}

One-dimensional spectra (1D) of the nebula were extracted by
summing up, without any weighting, all rows from the area of an
annulus with an outer radius of 10 arcsec centred on MN18 and the
central $\pm$3 arcsec excluded. Both low- and high-resolution 1D
spectra of the nebula are presented in Fig.\,\ref{fig:Neb}. The
emission lines detected in the spectra were measured using the
programs described in Kniazev et al. (2004). Table\,\ref{tab:int2}
lists the observed intensities of these lines normalized to
H$\beta$, F($\lambda$)/F(H$\beta$), the reddening-corrected line
intensity ratios, I($\lambda$)/I(H$\beta$), and the logarithmic
extinction coefficient, $C$(H$\beta$), which corresponds to
$E(B-V)$=2.24$\pm$0.15 mag. In Table\,\ref{tab:int2} we also give
the electron number density derived from the intensity ratio of
the [S\,{\sc ii}] $\lambda\lambda$6716, 6731 lines, $n_{\rm
e}$([S\,{\sc ii}]). The obtained value is comparable to densities
derived for equatorial rings of other bipolar nebulae around blue
supergiants and cLBVs (see Section\,\ref{sec:bip}). Both
$C$(H$\beta$) and $n_{\rm e}$([S\,{\sc ii}]) were calculated under
the assumption that $T_{\rm e}=10^4$ K. For lower temperatures
their values do not change much, e.g. $C$(H$\beta$)=3.27 and
$n_{\rm e}=510 \, {\rm cm}^{-3}$ if $T_{\rm e}$=7000\,K.

The electron number density could also be derived from the surface
brightness of the equatorial ring in the H$\alpha$ line, $S_{{\rm
H}\alpha}$, measured on the SHS image (cf. equation (4) in Frew et
al. 2014):
\begin{equation}
n_{\rm e}\approx 5 \, {\rm cm}^{-3} \left({l\over 1 \, {\rm
pc}}\right)^{-0.5}\left({S_{{\rm H}\alpha}\over 1 \, {\rm
R}}\right)^{0.5}e^{E(B-V)} \, ,
\end{equation}
where $l$ is the line of sight thickness of the ring and
1\,R$\equiv$1\,Rayleigh=5.66$\times10^{-18}$ erg cm$^{-2}$
s$^{-1}$ arcsec$^{-2}$ at H$\alpha$ (here we assume that $n_{\rm
e}$ is constant within the ring and $T_{\rm e}=10^4$ K).

Using equations\,(1) and (2) in Frew et al. (2014) and the flux
calibration factor of 7.2 counts pixel$^{-1}$ R$^{-1}$ from their
table\,1, and adopting the observed [N\,{\sc ii}] to H$\alpha$
line intensity ratio of 0.79 from Table\,\ref{tab:int2} (here
[N\,{\sc ii}] corresponds to the sum of the $\lambda$6548 and
$\lambda$6584 lines), we obtained the peak surface brightness of
the northeastern (brightest) half of the ring (corrected for the
contribution from the contaminant [N\,{\sc ii}] lines) of $S_{{\rm
H}\alpha}$$\approx$120 R. Using equation\,(1) with $E(B-V)$=1.97
mag and $l$$\approx$0.47 pc (the line of sight thickness of the
torus), one finds $n_{\rm e}$$\approx$570 ${\rm cm}^{-3}$, which
agrees well with $n_{\rm e}$([S\,{\sc ii}]) given in
Table\,\ref{tab:int2}.

To obtain an order of magnitude estimate of the mass of the
equatorial ring, $M_{\rm ring}$, we assume that the ring has a
toroidal geometry with the major and minor radii of $R$=6.5 arcsec
($\approx$0.17 pc) and $r$=4.5 arcsec ($\approx$0.12 pc),
respectively, and that its material is fully ionized and
homogeneous. In doing so, one finds $M_{\rm ring}$$\approx$$1 \,
\msun$. The actual value of $M_{\rm ring}$ could be higher if the
ionized gas encloses neutral and, correspondingly, more dense
material.

The [N\,{\sc ii}] and [S\,{\sc ii}] line intensities can be used
to estimate the nitrogen to sulphur abundance ratio, which is
almost independent of $n_{\rm e}$ and $T_{\rm e}$, provided that
$n_{\rm e}\leq 1000 \, {\rm cm}^{-3}$ and $T_{\rm e}\leq 10^4$ K.
In this case, the abundance ratio is given by (see Benvenuti,
D'Odorico \& Peimbert 1973 and references therein):
\begin{equation}
{N({\rm N}^+ )\over N({\rm S}^+ )} =3.61 {I(6584) \over
I(6716+6731)} \, . \label{eq:abud}
\end{equation}
Using equation\,(\ref{eq:abud}) and Table\,\ref{tab:int}, one
finds $N({\rm N}^+ )/N({\rm S}^+ )$$\approx$12.43$^{+6.05}
_{-4.04}$, which is $\approx$2.4$^{+1.2} _{-0.8}$ times larger
than the solar value of 5.12 (Asplund et al. 2009). Since the
sulfur abundance is not expected to change during stellar
evolution, one can argue that the high N to S ratio exists because
of an elevated nitrogen abundance in the line-emitting material.
Thus, using the solar N and S abundances from Asplund et al.
(2009), one finds a N abundance of the ring of
$\log$(N/H)+12=8.21.

\begin{figure}
\begin{center}
\includegraphics[angle=0,width=1.0\columnwidth,clip=]{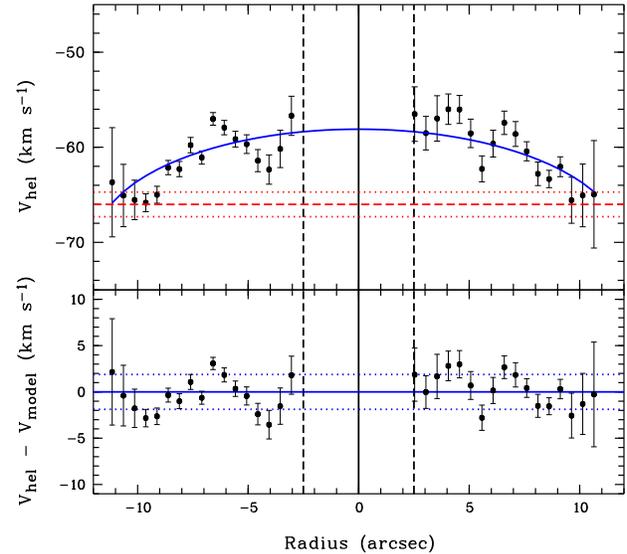}
\end{center}
\caption{{\it Upper panel:} H$\alpha$ heliocentric radial velocity
distribution along the slit. The solid vertical line corresponds
to the position of MN18, while the vertical dashed lines at
$\pm2.5\arcsec$ from the solid line mark the area where no nebular
emission was detected. The model fit is shown as a (blue) solid
line. The (red) dashed and dotted horizontal lines show,
respectively, the systemic velocity of the equatorial ring and its
error margins. {\it Bottom panel:} The difference between the
observed and model radial velocity distributions. The mean
residual velocity and its errors are shown with the (blue) solid
and dotted lines, respectively.} \label{fig:Vel_fit}
\end{figure}

\subsection{The position--velocity diagram and kinematics of the equatorial ring}
\label{sec:PV}

Fig.\,\ref{fig:Vel_fit} plots the distribution of the heliocentric
radial velocity of the H$\alpha$ line, $v_{\rm hel}$, along the
slit, calculated using the method and programs described in Zasov
et al. (2000). The systematic shifts originating from the RSS
flexure were excluded by taking into account the closest bright
night sky line. Finally, only those velocity measurements were
used which satisfy the criteria S/N\,$>3$ and $\sigma_v <6 \,
\kms$.

To estimate the systemic velocity of the equatorial ring, $v_{\rm
sys}$, and its expansion velocity, $v_{\rm exp}$, we fit the
observed distribution of $v_{\rm hel}$ with the model:
\begin{equation}
v_{\rm hel}=v_{\rm exp} \, \cos\alpha + v_{\rm sys} \, ,
\end{equation}
where $v_{\rm exp}$ and $v_{\rm sys}$ are free parameters, and
$\alpha$ is the angle between the line of sight and $v_{\rm exp}$
at each point of the ring. The result of the fit is shown as a
(blue) solid line in the upper panel of Fig.\,\ref{fig:Vel_fit}.
The (red) dashed and dotted horizontal lines show, respectively,
the derived $v_{\rm sys}$ of $-$66.0 $\kms$ and its error margins
($\pm$1.3 $\kms$). We also found $v_{\rm exp}=7.9$$\pm$1.2 $\kms$.
The difference between the observed and model velocity
distributions (the residual velocity hereafter) is shown in the
bottom panel of Fig.~\ref{fig:Vel_fit}, where the (blue) solid and
dotted lines indicate, respectively, the mean residual velocity of
$-0.05$ $\kms$ and $\pm$1$\sigma$ of the mean ($\pm$1.8 $\kms$).
The residual velocity distribution does not show any systematics,
which allows us to set an upper limit of 1.8 $\kms$ on the
rotational velocity of the ring.

The radius of the ring of $r$=0.29 pc together with $v_{\rm exp}$
yield the kinematic age of the ring of $t_{\rm kin}$=$r/v_{\rm
exp}$$\approx$$3.7\times10^4$ yr. Assuming that the equatorial
ring and bipolar lobes are the result of the same episode of
enhanced mass loss from MN18 (i.e. their $t_{\rm kin}$ are the
same; cf. Brandner et al. 1996b), one finds that the lobes should
expand at $\approx$$60 \, \kms$. Using $M_{\rm ring}$ from
Section\,\ref{sec:neb-phy} and $t_{\rm kin}$, one can also find an
order of magnitude ``kinematic" estimate of $\dot{M}$ during this
episode, i.e. $\dot{M}_{\rm kin}$$\sim$$M_{\rm ring}/t_{\rm
kin}$$\approx$$2.7$$\times$$10^{-5} \, \myr$, which is about two
orders of magnitude higher than the present-day values (see
Table\,\ref{tab:par}). The discrepancy between the mass-loss rates
would be somewhat smaller if the expansion of the equatorial ring
was faster initially. (Correspondingly, the expansion velocity of
the lobes would be smaller as well.) On the other hand, the
duration of the high mass loss episode could be shorter than
$t_{\rm kin}$, which would result in an even higher $\dot{M}_{\rm
kin}$.

\section{Discussion}
\label{sec:dis}

In Section\,\ref{sec:int}, we noted the existence of a class of
Galactic and extragalactic hourglass-like nebulae associated with
blue supergiants and mentioned seven known members of this class.
One of them -- the three-ring nebula associated with the blue
supergiant Sk$-$69$\degr$202 (the progenitor star of the
SN\,1987A) -- was formed shortly ($\sim$$10^4$\,yr) before the
star exploded as a supernova. Currently, one can observe how the
supernova blast wave interacts with the equatorial ring of the
nebula (Sonneborn et al. 1998; Sugerman et al 2002;
Gr\"{o}ningsson et al. 2008; Fransson et al. 2015). It is also
believed that LBVs could be immediate precursors of supernova
explosions (e.g. Kotak \& Vink 2006; Groh et al. 2013; Justham et
al. 2014). Thus, one can expect that some other massive stars with
bipolar nebulae will soon explode as supernovae and that the
appearance of their young (meaning from decades to centuries old)
diffuse supernova remnants would be affected by the dense material
of the nebulae. It is therefore worth to compare the derived
parameters of MN18 with those predicted by stellar evolutionary
models (e.g., Brott et al. 2011; Ekstr\"{o}m et al. 2012) to
determine the evolutionary status of this star and thereby to
understand whether or not it is close to explode as well. Knowing
the evolutionary status of MN18 would also allow us to constrain
the mechanism for the origin of the bipolar nebula around this
star, which in turn would help to better understand the origin of
bipolar nebulae around other massive stars.

\subsection{Evolutionary status of MN18}
\label{sec:stat}

The range of luminosities derived for MN18 in
Section\,\ref{sec:mod} corresponds to the initial mass, $M_{\rm
init}$, of this star of $\approx$30$\pm$$10 \, \msun$. Using the
grid of rotating main sequence stars by Brott et al. (2011), we
found that the main parameters of MN18 (temperature, luminosity,
gravity and CNO abundances) can be matched very well with models
with $M_{\rm init}$ in the above range, provided that the initial
rotational velocity of the star, $v_{\rm init}$, was of
$\sim$200--$400 \, \kms$. For example, the model of a $30 \,
\msun$ star with $v_{\rm init}=321 \, \kms$ predicts that after
$\approx$5.5 Myr the star would have the following parameters:
$T_{\rm eff}$$\approx$21.0$\pm$0.5\,kK,
$\log(L/\lsun)$$\approx$5.41, $R_*$$\approx$38$\pm$$2 \, \rsun$,
$\log g$$\approx$2.70$\pm$0.05 and a current rotational velocity
of $\approx$110$\pm$$10 \, \kms$. The CNO abundances of this star
would be $\approx$$6.3\times10^{-4}$, $1.6\times10^{-3}$ and
$3.4\times10^{-3}$, respectively. All these parameters are in good
agreement with those derived for MN18 (see Table\,\ref{tab:par}).
The models with low $v_{\rm init}$ also match very well the
$T_{\rm eff}$, $\log(L/\lsun)$ and $\log g$ values of MN18, but
their CNO abundances differ significantly. Specifically, the
nitrogen abundances predicted by these models are much lower than
what follows from our spectral modelling.

Similarly, we found that the rotating models by Ekstr\"{o}m et al.
(2012) also fit well the parameters of MN18 provided that this
star just left the main sequence, while in the case of a star
undergoing a blue loop evolution the CNO abundances predicted by
the models differ significantly from those of MN18. The
non-rotating models by Ekstr\"{o}m et al. (2012) do not fit the
CNO abundances of MN18 as well. Proceeding from this, one might
conclude that MN18 is a redward evolving star, which has only
recently become a blue supergiant (cf. Gvaramadze et al. 2014a).
This conclusion agrees with the suggestion by Lamers et al. (2001)
that LBV nebulae could be formed already at the end of the main
sequence phase and that their enhanced nitrogen abundances are due
to rotationally induced mixing in their underlying stars. Similar
suggestions were put forward by Smartt et al. (2002) and Hendry et
al. (2008) to explain the CNO abundances of the bipolar nebula
around Sher\,25.

The duplicity of massive stars provides another clue for
understanding the origin of enhanced N abundance in main sequence
stars because accretion of CNO-processed material from the mass
donor star or merger of the binary components can enhance the
surface N abundance of the mass gainer star or the merger product,
respectively, by a factor of several even in the absence of
rotational mixing\footnote{Models taking into account rotational
mixing in the mass gainers show that the N abundance can
additionally be enhanced by a factor of 3 (Langer et al. 2008).}
(Langer 2012; Glebbeek et al., 2013). As follows from the studies
by Mason et al. (2009), Sana \& Evans (2011) and Chini et al.
(2012), the majority of massive stars are members of close binary
systems. Furthermore, according to Sana et al. (2012), about 70
per cent of all massive stars are strongly affected by binary
interaction during their lifetimes through the mass transfer,
common envelope evolution or merger, and for more than half of
them the interaction takes place already during the main sequence
phase (see also de Mink et al. 2014). All of these types of binary
interaction can enhance the surface N abundance already during the
main sequence phase, and are also the likely culprits of the
production of substantial mass loss in the equatorial plane.

Thus, MN18 could be a mass gainer from a Case A or Case B system,
and as such it could either be rejuvenated or not. In the first
case there would be little difference to single star evolution,
while in the second one the star at its position in the
Hertzsprung-Russell diagram could either be core hydrogen or core
helium burning (Braun \& Langer 1995; Justham et al. 2014). The
latter possibility, however, is less likely because the N
abundance of the star would be much higher than what we found for
MN18.

To summarize, the above considerations suggest that MN18 is near
the end of the main sequence phase, which makes it unlikely that
this star will soon explode as a supernova.

\subsection{Origin of the bipolar nebulae}
\label{sec:bip}

Now we discuss the origin of bipolar nebulae around MN18 and other
massive stars. The rotational symmetry of these nebulae implies
that the stellar rotation plays a significant, if not a decisive,
role in their origin.

\subsubsection{Single star scenarios}
\label{sec:sin}

Lamers et al. (2001) proposed that at the end of the main sequence
the initially rapidly rotating stars may reach the limit of
hydrostatic stability ($\Omega$-limit; Langer 1997, 1998) in their
outer equatorial layers, where the joint action of the centrifugal
force and the radiation pressure balances the gravity (cf. Langer,
Gar\'{i}a-Segura \& Mac Low 1999). This may result in ejection of
a large amount of mass (up to several $\msun$) mainly in the
equatorial plane. The ejection process could be recurrent with
time intervals of $\sim10^3 -10^4$ yr, which are related to the
thermal time-scale of the stars. Correspondingly, the stellar wind
in between the ejections could be collimated in the polar
direction by the presence of the dense equatorial material.

To reach the $\Omega$-limit the star should be luminous enough to
arrive at or exceed its Eddington limit in the course of its
lifetime and/or be a rapid rotator. For single stars the first
condition could be satisfied during the advanced evolutionary
phases of very massive ($\ga$$50 \, \msun$) stars when their
masses are considerably reduced due to the extensive stellar wind.
Such stars can achieve the $\Omega$-limit even if they are slow
rotators (e.g. Langer \& Heger 1998). The resulting enhanced mass
loss of these stars is almost spherically symmetric and,
correspondingly, the circumstellar nebulae around them should be
almost spherical as well. In the case of very rapid rotators, the
$\Omega$-limit could be reached already during the main sequence.
This is conceivable even for those stars which never approach the
Eddington limit during their lifetime. Correspondingly, the mass
loss from these stars would be highly aspherical, with a slow and
dense wind at latitudes close to the equator and a fast and more
tenuous one in the polar direction. These stars would produce
elongated or bipolar nebulae.

Interestingly, the Brott et al.'s models predict that a
20$-$30$\,\msun$ star with $v_{\rm init}$$\approx$500$-$600$ \,
\kms$ can achieve the $\Omega$-limit by the end of the main
sequence, which results in a strong mass loss with $\dot{M}$ of
$\ga$$10^{-4} \, \myr$. But the higher $v_{\rm init}$, the
stronger the mixing, the longer the duration of the main sequence
phase and the bluer the star. Correspondingly, it takes longer for
the star's $T_{\rm eff}$ to decrease to 21 kK. The prolonged
rotational mixing in turn results in CNO abundances much different
from those derived from our spectral analysis.

The question on whether the surface velocity of single stars may
hit critical rotation during the main sequence evolution was
studied in detail by Ekstr\"{o}m et al. (2008). They computed
stellar models of various $M_{\rm init}$, $v_{\rm init}$ and
metallicities, and found that the $\Omega$-limit could be reached
by B stars with $v_{\rm init}$ in the upper tail of the initial
rotational velocity distribution. This could be responsible for
the origin of the Be phenomenon. Ekstr\"{o}m et al. (2008) also
found that much greater mass loss from more massive stars at high
(e.g. solar) metallicities prevents these stars from reaching the
critical rotation.

Heger \& Langer (1998) proposed a mechanism which may spin-up the
surface layers of rotating single stars to critical rotation after
they leave the red supergiant branch. It was suggested that this
mechanism may result in significant rotation of the progenitor
star of the SN1987A (which performed a blue loop just before the
supernova explosion) and thereby in the origin of the equatorial
ring around it. The CNO abundances of the material ejected by a
post-red supergiant star, however, would significantly differ from
those of MN18 (see Section\,\ref{sec:stat}).

Another possibility to increase mass loss rates of single stars is
based on the bi-stability in the radiation driven winds of
supergiants near the spectral type B1 (Pauldrach \& Puls 1990;
Lamers \& Pauldrach 1991), which is manifested in a drastic change
in the ionization states of Fe when $T_{\rm eff}$ drops a little
below some critical value ($\approx$21--25 kK; Vink, de Koter \&
Lamers 1999). This change leads to a factor of 2 decrease in
$v_\infty$, from $v_\infty/v_{\rm esc}$$\approx$2.6 for
supergiants of types earlier than B1 to $v_\infty/v_{\rm
esc}$$\approx$1.3 for B supergiants of later types (Lamers et al.
1995), and in a factor of several increase in $\dot{M}$ (Lamers et
al. 1995; Vink et al. 1999). The increase of $\dot{M}$ caused by
the bi-stability jump could be responsible for the ejection of
shells by the prototype LBV P\,Cygni (Pauldrach \& Puls 1990) and
the observed steep drop in the rotational velocities of massive
stars at $T_{\rm eff}$ of 22 kK (Vink et al. 2010).

More importantly, the enhanced mass loss caused by the
bi-stability mechanism could be asymmetric if the star is a rapid
rotator. In this case, the gravity darkening (the von Zeipel
effect) leads to a lower $T_{\rm eff}$ at the equator (e.g. Maeder
1999) and the bi-stability jump first occurs near the equator as
well. This in turn results in enhanced mass loss at low stellar
latitudes. Lamers \& Pauldrach (1991) and Lamers et al. (1995)
suggested that the bi-stability jump in the winds of rapidly
rotating early B-type supergiants could be responsible for the
origin of dense equatorial outflows (excretion disks) around B[e]
stars. The detailed study of this possibility by Pelupessy, Lamers
\& Vink (2000), however, showed that the bi-stability mechanism
can produce only about a four-fold increase in $\dot{M}$, which is
not enough to explain the equator to pole density contrast of a
factor of $\approx$100 observed for the winds of B[e] stars. It is
therefore doubtful that the bi-stability jump alone could explain
the about two order of magnitude difference between $\dot{M}$
derived from our spectral modelling of MN18 (see
Table\,\ref{tab:par}) and $\dot{M}$ required to produce the
equatorial ring around this star (see Section\,\ref{sec:PV}).

To conclude, although we cannot exclude the possibility that some
single massive stars can hit critical rotation in the course of
their evolution (resulting in massive equatorial outflows), we
believe that the origin of the equatorial ring around MN18 is more
likely because this star is or was a member of a close binary
system.

\subsubsection{Binary star scenarios}
\label{sec:bin}

The binary population synthesis modelling by de Mink et al. (2013)
suggests that about 20 per cent of massive main sequence stars
should be rapid ($>200 \, \kms$) rotators because of mass transfer
and mergers in close binary systems. About 2 per cent of these
stars are expected to have very high ($>400 \, \kms$) rotational
velocities (with the most extreme cases reaching $\approx$$600 \,
\kms$) of which a quarter are the merger remnants. de Mink et al.
(2013) noted that their predicted distribution of rotational
velocities of main sequence stars agrees well with the observed
one, which points to the possibility that mass transfer and
mergers are the main causes for the rapid rotation of massive
stars (cf. Packet 1981; Langer et al. 2003; Dufton et al. 2011).
It is tempting therefore to speculate that these two binary
interaction processes are the main causes for the origin of nearly
critically rotating stars and equatorial outflows from them.

The mass transfer in binary systems may not only spin-up the mass
gainers, but under proper conditions it could also result in the
origin of circumbinary disk-like structures by means of the common
envelope ejection (e.g. Morris 1981), mass loss through the second
Lagrange point (e.g. Livio, Salzman \& Shaviv 1979; Petrovic,
Langer \& van der Hucht 2005; Smith et al. 2011), and focusing of
the primary's wind towards the orbital plane by the gravitational
field of the companion star (e.g. Fabian \& Hansen 1979;
Mastrodemos \& Morris 1999). Similarly, the binary mergers can not
only produce the rapidly rotating stars, but also result in mass
ejection ($\sim$10 per cent of the binary mass; Suzuki et al.
2007), which is mainly concentrated in the binary's equatorial
plane. Moreover, the instant mass loss during the merger event may
bring the merger product to the Eddington limit, which in turn
would result in a copious mass loss by stellar wind.

If the origin of the equatorial ring around MN18 is indeed caused
by the equatorial mass loss from either a nearly critically
rotating mass gainer or a merger remnant, then the present
moderate rotational velocity of this star would imply that the
star has suffered significant angular momentum loss on a
time-scale of $\leq t_{\rm kin}$. The most efficient mechanisms
for the spin down of (single) massive stars is the magnetic
braking via a magnetized wind (Poe, Friend \& Cassinelli 1989;
ud-Doula, Owocki \& Townsend 2009), which for stellar magnetic
fields of $\sim$$10^3-10^4$\,G and sufficiently high $\dot{M}$ can
result in spin-down time-scales as short as $\sim$$10^4$\,yr. It
is not clear whether initially single stars can simultaneously
possess such high magnetic fields and be rapid rotators, but it is
believed that very strong large-scale magnetic fields can
intermittently be created because of the strong shear in merging
binary systems (Langer 2012; Wickramasinghe, Tout \& Ferrario
2014). Such magnetic fields can brake the rapid rotation of the
merger remnants through dynamical mass loss. Polarimetric
observations of MN18 and other stars with bipolar nebulae would be
of great importance to check whether they possess large-scale
magnetic fields of strength of $\sim$100--1000\,G, which are not
typical of the majority of massive main sequence stars (e.g. Wade
et al. 2014; Morel et al. 2015) and whose origin can be the
consequence of binary merger.

\begin{figure*}
\begin{center}
\includegraphics[width=1.5\columnwidth,angle=0]{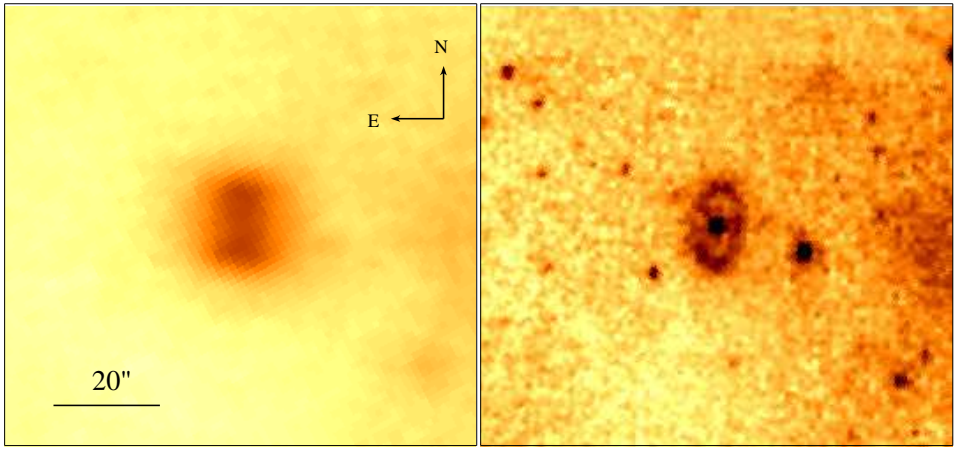}
\end{center}
\caption{{\it Spitzer} MIPS $24 \, \mu$m (left) and IRAC $8 \,
\mu$m images of a ring-like nebula around the possible blue
supergiant 2MASS\,J19281457+1716230.} \label{fig:mn109}
\end{figure*}

\begin{figure}
\begin{center}
\includegraphics[width=0.9\columnwidth,angle=270]{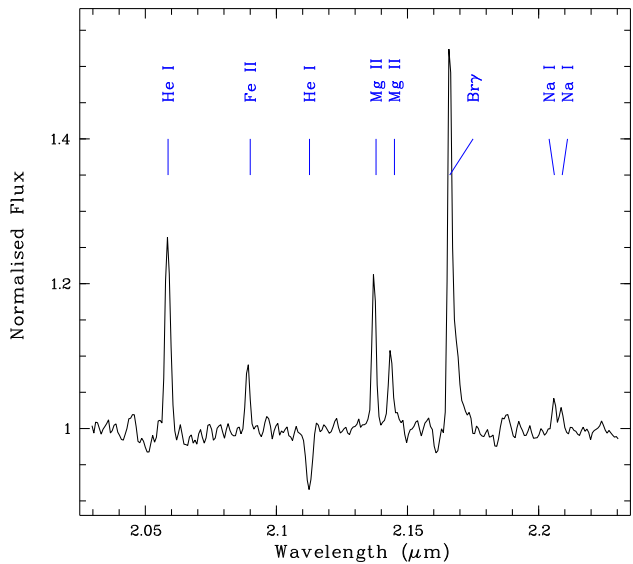}
\end{center}
\caption{Normalized VLT/ISAAC $K$-band spectrum of the cLBV MN56
with principal lines indicated.} \label{fig:mn56}
\end{figure}

In Section\,\ref{sec:PV}, we assumed that the kinematic ages of
the equatorial ring and bipolar lobes of the nebula around MN18
are the same (i.e. both components of the nebula were formed
simultaneously). But it is quite possible that the ring was formed
in the first instance (e.g. because of a dynamical mass ejection
caused by the binary merger) and that later on its dense material
collimated the stellar wind into bipolar lobes. If so, then pure
ring-like nebulae should exist whose central stars still have had
no time to blow-up pronounced bipolar lobes and which still might
be fast rotators. Indeed, several examples of such rings exist.
One of them, named [GKF2010]\,MN109 in the SIMBAD data base,
encircles the candidate blue supergiant 2MASS\,J19281457+1716230
(Phillips \& Ramos-Larios 2008). This ring is best visible at
8\,$\mu$m (see the right-hand panel of Fig.\,\ref{fig:mn109}),
while at 24\,$\mu$m there are no signatures of an extended bipolar
outflow. Another example, known as [GKF2010]\,MN56, is associated
with the cLBV MN56. In Fig.\,\ref{fig:mn56}, we present for the
first time the $K$-band spectrum of this star obtained with the
ISAAC spectrograph on the Very Large Telescope (VLT) during our
observing run in 2010\footnote{Full details of these observations
will be presented elsewhere, while some preliminary results could
be found in Stringfellow et al. (2012a,b).}. The spectrum shows
strong emission lines of H, He\,{\sc i}, Fe\,{\sc ii} and Mg\,{\sc
ii} and is almost identical to that of the cLBV HD\,316285 (see
fig.\,2 in Hillier et al. 1998). [GKF2010]\,MN56 also has a
clear-cut ring-like morphology with a ``horn" attached to the
south-east edge of the ring (see the right-hand panel of fig.\,4
in Paper\,I for the IRAC 5.8\,$\mu$m image of this structure). The
presence of the horn suggests that the stellar wind already starts
to be collimated into bipolar lobes. The saturated 24\,$\mu$m
image of [GKF2010]\,MN56 (see the left-hand panel of fig.\,4 in
Paper\,I) did not allow us to discern the horn, but it also did
not show signatures of a more extended bipolar outflow.
High-resolution spectroscopy of 2MASS\,J19281457+1716230 and MN56
would be desirable to determine the rotational velocities of these
two stars, as well as the evolutionary status of the former one.

Binary population studies suggest that the rate of massive binary
mergers in the Galaxy is $\sim$10--20 per cent of the Galactic
supernova rate, which corresponds to one per 200 yr (Langer 2012).
Assuming that all mergers result in the formation of ring-like or
bipolar nebulae and that the dispersion time-scale of these
nebulae is $\sim$$10^4$\,yr, one finds that our Galaxy should
contain about 100 such nebulae. The actual number of these nebulae
however should be much smaller because the majority ($\approx$80
per cent; Gies 1987) of massive stars reside in their parent star
clusters where the stellar winds and ionizing emission of
neighbouring hot massive stars could prevent the formation of
discernible circumstellar nebulae. This inference could be
supported by the fact that all massive stars with known
well-shaped (bipolar or ring-like) nebulae are located either at
the periphery of their parent star clusters or in the field, and
therefore are likely runaway stars (Gvaramadze et al. 2012a,b;
2014b).

\subsection{MN18 and other B supergiants with bipolar nebulae}
\label{sec:comp}

All but one (MN13) of seven hourglass-like nebulae around B
supergiants mentioned in Section\,\ref{sec:int} are visible (at
least partially) at optical wavelengths, and five of them were
studied by means of optical spectroscopy\footnote{There are no
reported spectroscopic observations of the equatorial belt
(visible in optical light; Marston \& McCollum 2008) of the
bipolar nebula around the cLBV MWC\,349A.}. In
Table\,\ref{tab:bip}, we compare the main parameters of the
central stars and equatorial rings of these five nebulae.

Table\,\ref{tab:bip} shows that the luminosities of the stars
associated with bipolar nebulae span a large range, from
$\log(L/\lsun$)= 4.7 to 5.8 (with the luminosity of MN18 being
close to the median value of $\approx$5.3). From this one can
conclude that the stellar luminosity (or $M_{\rm init}$) is not a
decisive factor in formation of bipolar nebulae, which is expected
if the nearly critical rotation is the main reason for the origin
of these nebulae. In this regard, it is worthy to note that some
PNe also possess the hourglass-like morphology, whose origin is
most likely due to binary interactions (e.g. Corradi \& Schwarz
1993; Sahai et al. 1999; De Marco 2009).

One can also see that $T_{\rm eff}$ values span quite a large
range as well, from 13 to 22 kK, which might be considered as an
indication that the bi-stability mechanism may not contribute to
the origin of at least some of the bipolar nebulae. This
inference, however, could be inconclusive because $T_{\rm eff}$ of
bona fide LBVs can vary in the same range on time scales as short
as several years (e.g. Stahl et al. 2001; Groh et al. 2009). One
cannot therefore exclude the possibility that $T_{\rm eff}$ of
Sk$-$69$\degr$202 and HD\,168625 were close to the bi-stability
limit during the formation of nebulae around these stars.

The axial symmetry of the bipolar nebulae suggests that the tilt
angles of their equatorial rings are equal to those of the
rotational axes of the central stars. This allows one to estimate
the rotational velocities for those stars whose projected
rotational velocities are known. Table\,\ref{tab:bip} shows that
none of these four stars are rapid rotators\footnote{There is no
information on the projected rotational velocity of the progenitor
star of the SN1987A.}. Thus, if the origin of the equatorial rings
is due to the nearly critical rotation of their underlying stars,
then the spin-down time-scales of these stars should be shorter
than $t_{\rm kin}$ of the equatorial rings, all of which are
$\sim$$10^4$ yr. It is reasonable to associate this drastic spin
down with a brief episode of very high $\dot{M}$ (e.g. Heger \&
Langer 1998), which is much higher than the current values of
$\dot{M}$ and perhaps even higher than the ``kinematic" ones,
$\dot{M}_{\rm kin}\sim M_{\rm ring}/t_{\rm kin}$.

Table\,\ref{tab:bip} also shows that the linear sizes of all
equatorial rings are almost the same: 0.2$-$0.3 pc. The largest
value of 0.3 pc was obtained for the equatorial ring around MN18
under the assumption that the distance to this object is 5.6 kpc.
At a shorter distance of $\approx$4.4 kpc (allowed by the error
margins of $d$; see also Section\,\ref{sec:clust}) the radius of
the ring would be similar to those of other equatorial rings. The
similarity in size of the rings produced by stars of different
$M_{\rm init}$ and evolutionary status, however, is most probably
a coincidence, because it is difficult to imagine any viable
mechanism for constraining it.

The number densities of all equatorial rings are also about the
same ($\sim 1000 \, {\rm cm}^{-3}$) except of the ring around
Sk$-$69$\degr$202, for which an order of magnitude higher value
was derived from the fading of the [O\,{\sc iii}] $\lambda$5007
emission line (Plait et al. 1995). The masses of the equatorial
rings range from $\approx$0.1 to $1 \, \msun$ with the lowest mass
belonging to the ring around Sk$-$69$\degr$202. This ring,
however, might be more massive if we see only its inner-faced
ionized part (e.g. Plait et al. 1995; cf. Fransson et al. 2015).

Finally, Table\,\ref{tab:bip} gives also the nitrogen abundances
(by number) of the equatorial rings. In the Sk$-$69$\degr$202 ring
the N abundance is a factor of 20 greater than the LMC in general
(7.14 dex; Russell \& Dopita 1992), which is consistent with the
fact that the material of this ring was ejected during the final
stage of the evolution of Sk$-$69$\degr$202. The Sher\,25 ring is
also greatly enriched with nitrogen as compared to the background
\hii region NGC\,3603 and the Sun, respectively, by factors of
$\approx$30 and 12 (Hendry et al. 2008). Although this
overabundance could be interpreted as an indication that Sher\,25
have evolved through the red supergiant stage (Brandner et al.
1997a,b), Hendry et al. (2008) argued that $M_{\rm init}$ of
Sher\,25 of $\sim$$50 \, \msun$ rather implies that this star did
not experience this stage at all. Instead, Hendry et al. (2008)
suggested that the enhanced N abundance is because of rotationally
induced mixing while the star was on the main sequence. Mass
accretion or a merger in a binary system might be responsible for
this enhancement as well (cf. Section\,\ref{sec:stat}). The
abundances of the rings produced by HD\,168625 and MN18 exceed the
solar abundance by a factor of few, while that of the [SBW2007]\,1
ring is subsolar. It is therefore likely that the underlying stars
of these three rings have not yet passed through the red
supergiant stage and that all these rings were formed at the end
of the main sequence stage or soon after it (cf. Lamers et al.
2001; Smith 2007; Smith et al. 2007; Gvaramadze et al. 2014a). The
largely different N abundances of the equatorial rings imply that
blue supergiants are capable to produce them in very different
evolutionary stages, from the main sequence stage up to the
pre-supernova stage. MN18 adds to the evidence that multiple ring
nebulae could be formed rather early during the stellar evolution.

\begin{table*}
\caption{Blue supergiants and cLBVs with hourglass-like
circumstellar nebulae.} \label{tab:bip}
\renewcommand{\footnoterule}{}
\begin{center}
\begin{minipage}{\textwidth}
\begin{tabular}{llcccc}
\hline
  & Sk$-$69$\degr$202 & Sher\,25 & HD\,168625 & [SBW2007]\,1 & MN18 \\
\hline
Spectral type & B3\,I$^{(1)}$ & B1.5\,Iab$^{(2)}$ & B6\,Iap$^{(3)}$ & B1\,Iab$^{(4)}$ & B1\,Ia \\
$\log(L/\lsun$) & $\approx$5$^{(5)}$ & 5.8$^{(6)}$ & 5.0$-$5.4$^{(7)}$ & 4.7$^{(8)}$ & 5.4 \\
$T_{\rm eff}$ (kK) & 16$^{(5)}$ & 22$^{(6)}$ & 12$-$15$^{(7)}$ & 21$^{(8)}$ & 21 \\
$\dot{M} (10^{-7} \, \myr)$ & 1.5$-$3$^{(9,10)}$ & 20$^{(6)}$ & 11$^{(7)}$ & 2$-$4$^{(8)}$ & 4.2$-$6.8 \\
$r$ (pc) & 0.2$^{(11)}$ & 0.2$^{(12)}$ & 0.2$^{(7)}$ & 0.2$^{(13)}$ & 0.3 \\
$n_{\rm e}$([S\,{\sc ii }]) (${\rm cm}^{-3}$) & $\sim$10\,000$^{(14,*)}$ & 500--1800$^{(15)}$ & $\approx$1000$^{(7)}$ & $\approx$500$^{(13)}$ & $\approx$600 \\
$M_{\rm ring}$ & 0.06$^{(16)}$ & 0.1$^{(15)}$ & 0.5$^{(7)}$ & 0.5--1.0$^{(8)}$ & 1 \\
$v\sin i (\kms$) & -- & 53 & 44 & 34 & 90 \\
$i$ ($\degr$) & 43$^{(17)}$ & 65$^{(12)}$ & 60$^{(18)}$ & 50$^{(13)}$ & 60 \\
$v_{\rm exp} (\kms$) & 10$^{(19)}$ & 30$^{(15)}$ & 20$^{(20)}$ & 19$^{(13)}$ & 8 \\
$t_{\rm kin}$ ($10^4$ yr) & 2 & 0.7 & 1 & 1 & 3.7 \\
$\dot{M}_{\rm kin}/\dot{M}$ & 10--20 & 7 & 45 & 250 & 40--60 \\
$\log$(N/H)+12 & 8.44$^{(16)}$ & 8.91$^{(6)}$ & 8.42$^{(7)}$ & 7.51$^{(13)}$ & 8.21 \\
\hline
\end{tabular}
\end{minipage}
\end{center}
References: (1) Walborn et al. 1989; (2) Moffat 1983; (3) Walborn
\& Fitzpatrick 2000; (4) Taylor et al. 2014; (5) Arnett et al.
1989; (6) Hendry et al. 2008; (7) Nota et al. 1996; (8) Smith et
al. 2013; (9) Blondin \& Lundqvist 1993; (10) Martin \& Arnett
1995; (11) Panagia et al. 1991; (12) Brandner et al. 1997a; (13)
Smith et al. 2007; (14) Plait et al. 1995; (15) Brandner et al.
1997b; (16) Mattila et al. 2010; (17) Jakobsen et al. 1991; (18)
O'Hara 2003; (19) Crotts \& Heathcote 1991; (20) Hutsemekers et
al.
1994. \\
{\it Note:} $^*$Based on the fading of the [O\,{\sc iii}]
$\lambda$5007 emission line.
\end{table*}

\subsection{MN18 and Lynga\,3}
\label{sec:clust}

\begin{table*}
\caption{Peculiar transverse (in Galactic coordinates) and radial
velocities, and the total space velocity of MN18 for two adopted
proper motion measurements and three distances (see the text for
details).} \label{tab:prop}
\renewcommand{\footnoterule}{}
\begin{center}
\begin{minipage}{\textwidth}
\begin{tabular}{ccccccc}
\hline $d$ & $\mu _\alpha \cos \delta$ & $\mu _\delta$ & $v_l$ & $v_b$ & $v_{\rm r}$ & $v_\ast$ \\
(kpc) & (mas ${\rm yr}^{-1}$) & (mas ${\rm yr}^{-1}$) & ($\kms$) & ($\kms$) & ($\kms$) & ($\kms$) \\
\hline
4.4 & $-10.3$$\pm$2.0$^a$ & $-7.6$$\pm$2.0$^a$ & $-144.3$$\pm$42.1 & $-12.6$$\pm$41.9 & 9.6$\pm$1.3 & 145.2$\pm$42.0 \\
4.4 & $-12.3$$\pm$2.5$^b$ & $-10.3$$\pm$2.5$^b$ & $-209.7$$\pm$52.1 & $-39.5$$\pm$52.1 & 9.6$\pm$1.3 & 213.6$\pm$52.0 \\
& & & & & & \\
5.6 & $-10.3$$\pm$2.0$^a$ & $-7.6$$\pm$2.0$^a$ & $-164.4$$\pm$53.6 & $-18.2$$\pm$53.4 & 22.1$\pm$1.3 & 166.9$\pm$53.1 \\
5.6 & $-12.3$$\pm$2.5$^b$ & $-10.3$$\pm$2.5$^b$ & $-247.8$$\pm$66.4 & $-52.4$$\pm$66.4 & 22.1$\pm$1.3 & 254.2$\pm$66.2 \\
& & & & & & \\
7.1 & $-10.3$$\pm$2.0$^a$ & $-7.6$$\pm$2.0$^a$ & $-184.3$$\pm$68.0 & $-25.3$$\pm$67.6 & 20.3$\pm$1.3 & 187.1$\pm$67.6 \\
7.1 & $-12.3$$\pm$2.5$^b$ & $-10.3$$\pm$2.5$^b$ & $-290.0$$\pm$84.1 & $-68.7$$\pm$84.1 & 20.3$\pm$1.3 & 298.7$\pm$83.9 \\
\hline
\end{tabular}
\end{minipage}
\end{center}
{\it Note:} $^a$SPM\,4.0 (Girard et al. 2011); $^b$UCAC4
(Zacharias et al. 2013).
\end{table*}

Is Section\,\ref{sec:neb}, we mentioned that Lynga (1964a) listed
MN18 among the members of the newly identified open star cluster
Lynga\,3. Lynga (1964b) noted that the cluster is strongly
reddened, which did not allow him to estimate the interstellar
extinction and the distance to the cluster.

A literature search revealed three subsequent studies of Lynga\,3,
based either on infrared or optical photometry. Carraro et al.
(2006) constructed a colour-magnitude diagram of what they assumed
to be a cluster using their own CCD $BVI$ photometry and did not
find any significant difference with respect to a comparison
field. This led them to conclude that ``Lynga\,3 is not a cluster,
but an overdensity of a few bright stars probably located inside
the Carina arm". It turns out, however, that they confused the
1950 epoch coordinates with the 2000 epoch ones, which resulted in
more than $0\fdg5$ offset between the group of stars they analyzed
and the centre of the cluster given in Lynga (1964a,b). Moreover,
the inspection of the finding chart of Lynga\,3 given in fig.\,13
of Lynga (1964a) revealed that the geometric centre of the
putative cluster is offset by about $5\farcm5$ from the cluster's
position listed in Lynga (1964a,b; see tables\,V in these papers).
This explains why Carraro et al. (2006) came to the conclusion
that Lynga\,3 is not a true cluster.

More recently, Lynga\,3 was studied by Kharchenko et al. (2013)
and Dias et al. (2014), who used photometric and proper motion
measurements from different catalogues to evaluate the membership
probabilities for individual stars in the cluster region and to
derive the distance to and age of the cluster. Although the
coordinates of the cluster given in these two references are in a
quite good agreement with each other (as well as with the position
of the cluster's geometric centre implied by the finding chart in
Lynga 1964a), the obtained distance estimates, 4.2 kpc (Kharchenko
et al. 2013) and 1.4 kpc (Dias et al. 2014), differ from each
other drastically. Moreover, the age estimates of $\log t$=8.92
(Kharchenko et al. 2013) and 8.55 (Dias et al. 2014) are much
older than the age of MN18 of $\log t$$\approx$7 (see
Section\,\ref{sec:stat}). Since this huge age difference cannot be
explained by rejuvenation of MN18 because of mass transfer or
merger in a binary system, it is reasonable to assume that MN18
and the putative star cluster are projected by chance along the
same line-of-sight. This raises the question where the parent
cluster of MN18 is?

A possible answer to this question is that MN18 is a runaway star,
which was ejected and arrived at its current position from a
nearby massive star cluster (cf. Gvaramadze, Pflamm-Altenburg \&
Kroupa 2011; Gvaramadze et al. 2012a). The runaway status of this
star is suggested by proper motion measurements given in the
SPM\,4.0 (Girard et al. 2011) and UCAC4 (Zacharias et al. 2013)
catalogues (see Table\,\ref{tab:prop}). Using these measurements,
$v_{\rm sys}$ derived in Section\,\ref{sec:PV}, the Solar
galactocentric distance $R_0 = 8.0$ kpc and the circular Galactic
rotation velocity $\Theta _0 =240 \, \kms$ (Reid et al. 2009), and
the solar peculiar motion
$(U_{\odot},V_{\odot},W_{\odot})=(11.1,12.2,7.3) \, \kms$
(Sch\"onrich, Binney \& Dehnen 2010), we calculated the peculiar
transverse (in Galactic coordinates) velocity, $v_{\rm l}$ and
$v_{\rm b}$, the peculiar radial velocity, $v_{\rm r}$, and the
total space velocity, $v_\ast$, of MN18 (see
Table\,\ref{tab:prop}). To take into account uncertainties in the
distance estimate, we adopted three values of the distance:
$d$=4.4, 5.6 and 7.1 kpc (see Section\,\ref{sec:mod}). For the
error calculation, only the errors of the proper motion and the
systemic velocity measurements were considered.

\begin{figure*}
\begin{center}
\includegraphics[width=1.8\columnwidth,angle=0]{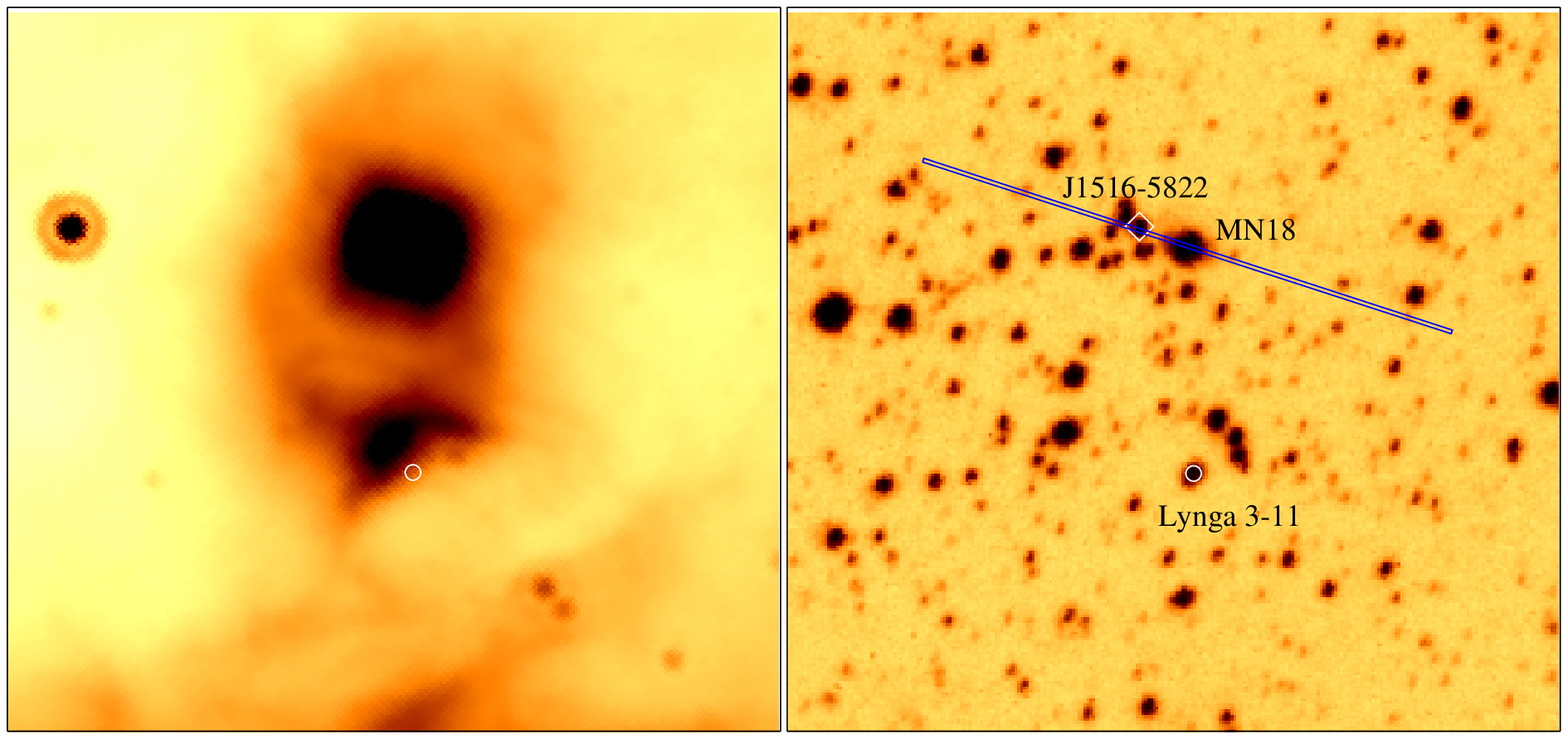}
\end{center}
\caption{{\it Spitzer} MIPS $24 \, \mu$m (left) and SHS
H$\alpha$+[N\,{\sc ii}] images of a field around MN18 with
positions of the newly-identified late O-/early B-type star
2MASS\,J15164297$-$5822197 and star\,11 from Lynga (1964a)
indicated by a diamond and circle, respectively. The orientation
of the RSS slit (PA=72$\degr$) is shown by a (blue) rectangle. See
text for details.} \label{fig:neb-3}
\end{figure*}

\begin{figure*}
\begin{center}
\includegraphics[width=11cm,angle=270,clip=]{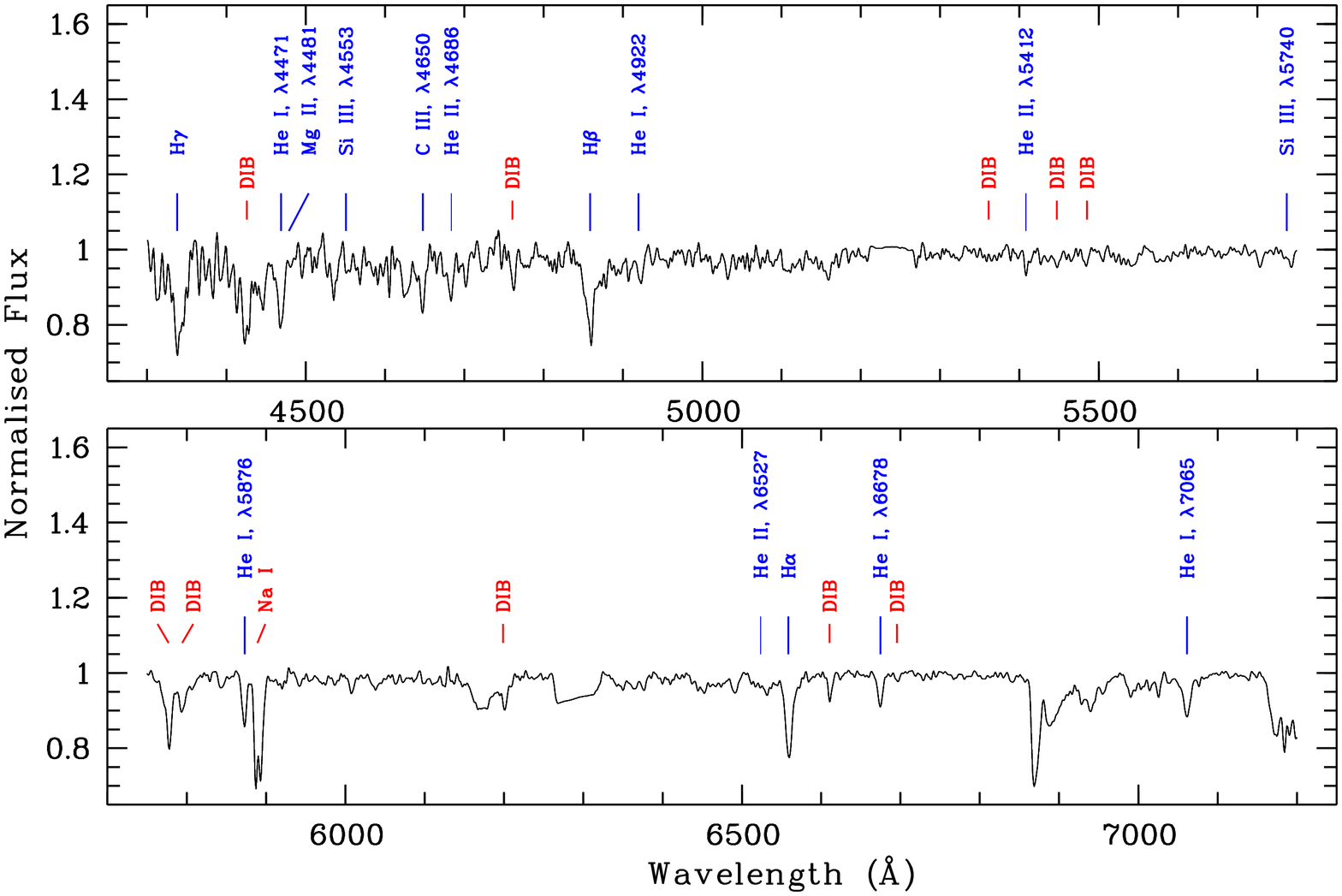}
\end{center}
\caption{Normalized spectrum of 2MASS\,J15164297$-$5822197
obtained with the SALT on 2012 March 2 with the principal lines
and most prominent DIBs indicated.} \label{fig:spec-l}
\end{figure*}

From Table\,\ref{tab:prop} it follows that if the SPM\,4.0 and
UCAC4 proper motion measurements are correct then MN18 is a
high-velocity runaway star, moving almost along the Galactic plane
and the plane of sky. In this case, the parent cluster of MN18
would be at $\sim$15$\degr$--25$\degr$ from the current position
of the star (i.e. at $l$$\approx$336$\degr$--346$\degr$), provided
that it was ejected at the very beginning of the cluster's
dynamical evolution (i.e. $\sim$5--10 Myr ago). It should be
noted, however, that although the high space velocity derived for
MN18 is not unusual for single runaway OB stars, it would be
unrealistically high if this star have been ejected as a binary
(cf. Section\,\ref{sec:bip}). Moreover, the ram pressure of the
interstellar medium due to the stellar space velocity would affect
the shape of the bipolar nebula, sweeping back the polar lobes,
which is definitely not the case for the nebula around MN18. One
cannot therefore exclude the possibility that the proper motion
measurements for MN18 are erroneous (e.g. because of the presence
of the equatorial ring) and that this star was actually born not
far from its current position on the sky. There is some evidence
in support of this possibility. Before discussing this evidence,
we note that the small peculiar radial velocity of MN18 implies
that the distance to this star did not change appreciably during
its lifetime. Since the peculiar transverse velocity of MN18 is
unrealistically high, and massive stars did not form in the
interarm regions, we conclude that the most likely birthplace of
MN18 is located in the Crux-Scutum Arm, i.e. at $d$$\sim$4 kpc.

In Section\,\ref{sec:neb-phy}, we noted that the RSS slit crosses
two other stars on both sides of MN18 (see also
Fig.\,\ref{fig:neb-3}). The normalized spectrum of the brightest
of these stars, located at RA(2000)=$15^{\rm h} 16^{\rm m}
42\fs97$, Dec.(2000)=$-$$58\degr 22\arcmin 19\farcs7$ and
indicated in Fig.\,\ref{fig:neb-3} by a diamond, is shown in
Fig.\,\ref{fig:spec-l}. In the following, we will use for this
star its 2MASS name -- 2MASS\,J15164297$-$5822197, or
J1516$-$5822, in short. Like in the case of MN18, the spectrum of
J1516$-$5822 is dominated by absorption lines of H and He\,{\sc
i}. But additionally, it also shows weak absorption lines of
He\,{\sc ii} at $\lambda\lambda$4686, 5412 and 6527, which implies
that J1516$-$5822 is of a spectral type earlier than B1 (Walborn
\& Fitzpatrick 1990), i.e. hotter than MN18. The Mg\,{\sc ii}
$\lambda$4481 line is weak, which is typical of late O/B0 stars.
The low quality of the blue part of the spectrum precludes us from
using traditional classification criteria. Instead, we used the
EW(H$\gamma$)--absolute magnitude calibration by Balona \&
Crampton (1974), which for EW(H$\gamma$)=3.14$\pm$0.20 \AA \,
implies that the spectral type of J1516$-$5822 is either O9\,V,
B0--1\,III or B5--7\,Ib. The B5--7\,Ib classification could be
rejected because of the presence of the He\,{\sc ii} lines in the
spectrum. Thus, we tentatively classify J1516$-$5822 as a
O9\,V/B0--1\,III star. EWs, FWHMs and RVs of main lines in the
spectrum of J1516$-$5822 are summarized in
Table\,\ref{tab:Intens}.

Detection of a massive and therefore very rare star at only
$\approx$17$\arcsec$ from MN18 suggests that both stars could be
members of the same star cluster. Assuming that J1516$-$5822 is an
O9\,V star and using its 2MASS magnitudes (J,H,K$_{\rm
s}$)=(11.35,10.85,10.63) (Cutri et al. 2003) and the photometric
calibration of optical and infrared magnitudes for Galactic O
stars by Martins \& Plez (2006), we derived a $K$-band extinction
towards J1516$-$5822 of $A_K=0.59$ mag and a distance modulus of
13.36 mag, which corresponds to the distance of 4.7 kpc. The
distance to J1516$-$5822 would be of $\approx$7 kpc if the
spectral type of this star is B0--1\,III. These estimates are
consistent with the distance estimate of $5.6^{+1.5}_{-1.2}$ for
MN18 (see Section\,\ref{sec:mod}), which supports the supposition
that both stars might be members of the same cluster. Note that
the mean RV of lines in the spectrum of J1516$-$5822 of $145 \,
\kms$ is more than twice higher than $v_{\rm sys}$ of MN18.
Although this difference might indicate that the two stars are
unrelated to each other and are simply projected by chance along
the same line-of-sight, it could also be understood if
J1516$-$5822 is a close massive binary. Note also that the
SPM\,4.0 and UCAC4 proper motion measurements for this star ($\mu
_\alpha \cos \delta =-28.42\pm2.89, \, \mu _\delta =45.47\pm2.92$
and $\mu _\alpha \cos \delta =-36.0\pm3.5, \, \mu _\delta
=56.6\pm3.4$, respectively) differ significantly from those of
MN18. These measurements, however, seem to be erroneous because
they imply a huge space velocity of $\sim$1000--2000 $\kms$ for
any reasonable distance to the star. To conclude, the existing
data did not allow us to prove or disprove the physical
association between MN18 and J1516$-$5822.

\begin{table}
\centering{ \caption{EWs, FWHMs and RVs of main lines in the
spectrum of 2MASS\,J15164297$-$5822197.} \label{tab:Intens}
\begin{tabular}{lccc} \hline
\rule{0pt}{10pt} $\lambda_{0}$(\AA) Ion  & EW($\lambda$) &
FWHM($\lambda$) & RV \\
& (\AA) & (\AA) & ($\kms$) \\
\hline
4340\ H$\gamma$\  &  3.14$\pm$0.20 & 15.03$\pm$2.53 &  -96$\pm$16    \\
4471\ He\ {\sc i}\      &  1.00$\pm$0.04 &  9.58$\pm$0.40 &  -158$\pm$9    \\
4922\ He\ {\sc i}\      &  0.25$\pm$0.04 &  4.50$\pm$0.51 &  -163$\pm$8    \\
5876\ He\ {\sc i}\      &  0.87$\pm$0.04 &  6.73$\pm$0.27 &  -141$\pm$6    \\
6563\ H$\alpha$\    &  1.74$\pm$0.05 & 10.62$\pm$0.24 &  -146$\pm$6    \\
6678\ He\ {\sc i}\      &  0.45$\pm$0.03 &  7.03$\pm$0.17 &  -150$\pm$6    \\
7065\ He\ {\sc i}\      &  0.71$\pm$0.03 & 11.16$\pm$0.29 &  -161$\pm$5    \\
\hline
\end{tabular}
  }
\end{table}

Finally, we discuss the bright arc-like feature attached to the
southwestern edge of the southeast lobe of the bipolar nebula
around MN18 (see Section\,\ref{sec:neb} and Figs\,\ref{fig:neb}
and \ref{fig:neb-3}). Fig.\,\ref{fig:neb-3} shows that there is a
star within the arc and near its apex. This star, located at
RA(J2000)=$15^{\rm h} 16^{\rm m} 40\fs76$, Dec.(J2000)=$-58\degr
23\arcmin 38\farcs7$, is listed in table\,V of Lynga (1964a) as a
member of Lynga\,3 under the number 11 (Lynga\,3-11 hereafter). We
speculate that the arc might be created because of interaction
between the wind of Lynga\,3-11 and the nebula (cf. Mauerhan et
al. 2010; Burgemeister et al. 2013). The enhanced brightness of
the arc at the place of possible contact with the nebula supports
this possibility. If the interaction indeed takes place, then
Lynga\,3-11 should be a source of strong wind, i.e. an OB star. In
the absence of spectroscopic observations of Lynga\,3-11, we
estimate its spectral type using its 2MASS magnitudes (J,H,K$_{\rm
s}$)=(11.07,10.59,10.34) and a photometric calibration for
Galactic O stars by Martins \& Plez (2006), and assuming that this
star is located at the same distance as MN18 (i.e. at
$d=5.6^{+1.5}_{-1.2}$ kpc). In doing so, we found that Lynga\,3-11
should be of spectral type of O9.5\,V ($d$=4.4 kpc), O6\,V
($d$=5.6 kpc) or 04\,V/O7.5\,III ($d$=7.1 kpc). A spectroscopic
study of this and other stars listed in Lynga (1964a) is necessary
to check whether or not they form the parent cluster of MN18.

\section{Summary}

In this Paper, we presented the results of optical spectroscopic
observations of the blue supergiant MN18 and the equatorial ring
of its hourglass-like nebula. MN18 was classified as B1\,Ia and
its spectrum was modelled using the radiative transfer code {\sc
cmfgen}. We found the stellar effective temperature of
21.1$\pm$1.0 kK and the surface gravity of $\log g$=2.75$\pm$0.20.
Adopting an absolute visual magnitude of $M_V$=$-$6.8$\pm$0.5
(typical of B1 supergiants), MN18 has a luminosity of $\log
L/\lsun$$\approx$5.42$\pm$0.3 and resides at a distance of
$\approx$5.6$^{+1.5} _{-1.2}$ kpc. It was shown that the nitrogen
abundance of the stellar wind is enhanced by $\approx$4 times due
to the CNO cycle and that the rotational velocity of MN18 is
$\approx$$100 \, \kms$. The H$\alpha$ line in the spectrum of MN18
changes from emission to absorption on a yearly time scale and
shows a noticeable variability on a daily time scale.
Correspondingly, the mass-loss rate estimates based on
synthesizing the H$\alpha$ line show a range of values from
$\approx$2.8 to $4.5\times10^{-7} \, \myr$. Our photometric
observations showed that the $B,V$ and $I_{\rm c}$ magnitudes of
MN18 changed by less than $\approx$0.1 mag during the last five
years, while its $B-V$ and $B-I_{\rm c}$ colours remained almost
constant.

Analysis of the nebular spectrum revealed a factor of $\approx$2
overabundance of nitrogen, which implies that the equatorial ring
is composed of processed material. The mass of the ring was found
to be of $\approx$$1 \, \msun$, and its systemic and expansion
velocities were estimated to be of $-$66.0$\pm$$1.3 \, \kms$ and
7.9$\pm$$1.2 \, \kms$, respectively. The latter estimate implies
the kinematic age of the equatorial ring of
$\approx$$3.7\times10^4$ yr.

The derived parameters of MN18 were compared with those predicted
by stellar evolutionary models and a conclusion was drawn that
this star is in the pre-red supergiant phase of its evolution.
This conclusion agrees with previous suggestions that
circumstellar nebulae around massive stars could be formed already
near the end of the main sequence phase, and allowed us to suggest
that the origin of the equatorial ring around MN18 is unlikely to
be explained within the framework of single star evolution. Based
on the observed high percentage of binaries among massive stars
and the possibility that massive stars did not born as rapid
rotators, but became such because of mass transfer or mergers in
binary systems, we suggested that these two binary interaction
processes might be the main causes for the origin of nearly
critically rotating stars and equatorial outflows from them.

The parameters of MN18 and its equatorial ring were also compared
with those of four other blue supergiants with hourglass-like
nebulae in the Milky Way (Sher\,25, HD\,168625, [SBW2007]\,1) and
the Large Magellanic Cloud (Sk$-$69$\degr$202). In particular, the
comparison revealed that the luminosities of these stars span a
wide range, from $\log(L/\lsun$)= 4.7 to 5.8, which implies that
the stellar luminosity is not a decisive factor in formation of
equatorial rings. Also, the nitrogen abundances of the equatorial
rings around MN18, HD\,168625 and [SBW2007]\,1 are consistent with
the possibility that these rings were formed at the end of the
main sequence stage or soon after it.

We found that MN18 is among a group of 25 stars suggested in the
literature as being members of the star cluster Lynga\,3. The
reported age estimates for the cluster, however, about two orders
of magnitude larger than the age of MN18, which raises doubts in
their association. On the other hand, our spectroscopic
observations led to a serendipitous discovery of a late O-/early
B-type star at an angular separation of only $\approx$17 arcsec
from MN18. The close proximity of the two massive and therefore
very rare stars suggests that both could be members of the same
young star cluster. The distance estimates for these two stars
support this suggestion. Moreover, we found an indication that
another member of the putative cluster Lynga\,3 is a massive star
as well, which also points to the possibility that Lynga\,3 is a
young star cluster. Spectroscopic follow-ups of this and other
stars around MN18 are necessary to prove whether Lynga\,3 is the
parent cluster to MN18.

\section{Acknowledgements}
The observations reported in this paper were obtained with the
Southern African Large Telescope (SALT) and the 2.2m MPG telescope
at La Silla (ESO), which is operated by MPIA Heidelberg and MPE
Garching. We thank John Hillier for {\sc cmfgen}, Wolf-Rainer
Hamann for {\sc wrplot}, and Helmut Steinle (MPE) for executing
the MN18 FEROS observations. VVG and LNB acknowledge the Russian
Science Foundation grants 14-12-01096 and 14-22-00041,
respectively. AYK acknowledges support from the National Research
Foundation (NRF) of South Africa. This research has made use of
the NASA/IPAC Infrared Science Archive, which is operated by the
Jet Propulsion Laboratory, California Institute of Technology,
under contract with the National Aeronautics and Space
Administration, the SIMBAD data base and the VizieR catalogue
access tool, both operated at CDS, Strasbourg, France, and the
WEBDA data base, operated at the Department of Theoretical Physics
and Astrophysics of the Masaryk University.

\end{document}